\documentclass[extra,mreferee]{gji}
\usepackage{graphicx}
\usepackage{amsmath}
\usepackage{amssymb}
\usepackage{lineno}
\usepackage{multirow}
\usepackage[section]{placeins}

\title{
  Direct current resistivity with steel-cased wells
}

\author[Lindsey J. Heagy, Douglas W. Oldenburg]{
  Lindsey J. Heagy$^1$ and Douglas W. Oldenburg$^1$\\
  $^1$ Geophysical Inversion Facility, University of British Columbia, Vancouver, BC, Canada, V6T 1Z4
  \\ \quad email: lheagy@eos.ubc.ca
}

\begin{document}

\label{firstpage}

\maketitle

\begin{summary}

The work in this paper is motivated by the increasing use of electrical and electromagnetic methods in geoscience problems where steel-cased wells are present. Applications of interest include monitoring carbon capture and storage and hydraulic fracturing operations, as well as detecting flaws or breaks in degrading steel-casings -- such wells pose serious environmental hazards. The general principles of electrical methods with steel-cased wells are understood, and several authors have demonstrated that the presence of steel-cased wells can be beneficial for detecting signal due to targets at depth. However, the success of a DC resistivity survey lies in the details. Secondary signals might only be a few percent of the primary signal. In designing a survey, the geometry of the source and receivers, and whether the source is at the top of the casing, inside of it, or beneath the casing will impact measured responses. Also  the physical properties and geometry of the background geology, target, and casing will have a large impact on the measured data. Because of the small values of the diagnostic signals, it is important to understand the detailed physics of the problem and also to be able to carry out accurate simulations. This latter task is computationally challenging because of the extreme geometry of the wells, which extend kilometers in depth but have millimeter variations in the radial direction, and the extreme variation in the electrical conductivity (typically 5-7 orders of magnitude between the casing and the background geology).

In this paper, we adopt a cylindrical discretization for numerical simulations to investigate three important aspects of DC resistivity in settings with steel-cased wells. (1) We examine the feasibility of using a surface-based DC resistivity survey for diagnosing impairments along a well in a casing integrity experiment. This parameter study demonstrates the impact of the background conductivity, the conductivity of the casing, the depth of the flaw, and the proportion of the casing circumference that is compromised, on amplitude of the secondary electric fields measured at the surface. (2) Next, we consider elements of survey design for exciting a conductive or resistive target at depth. We show that conductive targets generate stronger secondary responses than resistive targets, and that having an electrical connection between the target and well can significantly increase the measured secondary responses. (3) Finally, we examine common strategies for approximating the fine-scale structure of a steel cased well with a coarse-scale representation to reduce computational load. We show that for DC resistivity experiments, the product of the conductivity and the cross-sectional area of the casing is the important quantity for controlling the distribution of currents and charges along its length.

To promote insight into the physics, we present results by plotting the currents, charges, and electric fields in each of the scenarios examined. All of the examples shown in this paper are built on open-source software and are available as Jupyter notebooks.

\end{summary}



\section{Introduction}

Subsurface resistivity can be a valuable part of a geologic interpretation, whether that be identifying lithologic units, characterizing changes within a reservoir, or imaging subsurface injections associated with carbon capture and storage or hydraulic fracturing. In many of these settings, steel-cased wellbores are present. Steel has a significant electrical conductivity, which is generally six or more orders of magnitude larger that of the surrounding of the geologic formation. Clearly, such a large contrast is important to consider when conducting a direct current (DC) resistivity survey. On one-hand, the role of the steel casing may be viewed as ``distortion'' which complicates the signals of interest \citep{Wait1983, Holladay1984, Johnston1987}. In other scenarios, a wellbore may be beneficial in that it can serve as an ``extended electrode'' so that current-injection and sampling of the resultant electrical potentials can take place beneath near surface heterogeneities \citep{Ramirez1996, Rucker2010, Rucker2012, Ronczka2015} or so that currents injected at the surface can reach significant depths \citep{Schenkel1994, Weiss2016, hoversten2017borehole}. The use of casings as extended electrodes extends back several decades. \cite{Sill1978} used the well casing as a buried electrode for their mise-\`a-la-masse experiment at the Roosevelt Hot Springs geothermal field in Utah, as did \cite{Kauahikaua1980} for their mise-\`a-la-masse mapping of a high temperature geothermal reservoir in Hawaii. \cite{Sill1983} used the well as a source to monitor an injection test at Raft River, Idaho to determine if measurable changes that might indicate the direction of fluid flow could be observed. \cite{Rocroi1985} delineated a known resistive hydrocarbon deposit in the USSR by injecting current into two cased wells. More recently, applications for hydraulic fracturing, enhanced oil recovery and carbon capture and storage have been of much interest \citep{Commer2015, Tietze2015, Um2015, Weiss2016, hoversten2017borehole}. There has also been an increase in interest in examining the use of electrical or electromagnetic methods deployed on the surface to non-invasively look for flaws or breaks in the casing. \cite{Wilt2018} introduces the idea of using electrical or electromagnetic methods for casing integrity which is further expanded upon in \cite{Wilt2018a}. They show that low-frequency electromagnetic methods are sensitive to variations in wellbore length and demonstrate that their numerical simulations agree with field data collected over two different wellbores at the Containment and Monitoring Institute (CaMI) field site in southern Alberta, Canada. This work provides motivation for further delving into the physics and assessing under which circumstances we can expect to detect a flaw along a wellbore using electrical or electromagnetic methods.

To build a physical understanding of electrical and electromagnetic methods in settings where steel-cased wells are present, there are several areas to be investigated. First, the significant conductivity of the steel will impact the behavior of the charges, currents, and electric fields. This is true at the electrostatic limit, relevant to DC resistivity surveys, as well as when the source fields are time-varying, as in electromagnetic (EM) surveys. When considering EM surveys, induction effects also influence the responses, and magnetic fields and fluxes become relevant, meaning that the magnetic permeability of the steel then introduces further complexity into the signals we measure. This paper is concerned with the first set of physical phenomena: understanding the physics of steel casings at DC.

Much of the initial theory and understanding of the behavior of electric fields, currents, and charges, was developed in the context of well-logging. \cite{Kaufman1990} and \cite{Kaufman1993} provide a theoretical basis for our understanding; the first paper derives an analytical solution for a DC experiment where an electrode is positioned along the axis of an infinite length well, and discusses where charges accumulate and how currents leak into the surrounding formation. From this, \cite{Kaufman1990} shows that by measuring the second derivative of the electric potential, information about the formation resistivity can be obtained. The second paper extends the analysis for finite length wells. \cite{Schenkel1990, Schenkel1991, Schenkel1994} pioneered numerical work analyzing the influence of steel-cased wells on geophysical data using an integral equation approach for solving the DC resistivity problem. They expand upon the logging-through-casing application and discuss limitations of the transmission line solution presented in \cite{Kaufman1990} for this application. They also explored the feasibility of cross-hole and borehole-to-surface surveys where one electrode is placed within, or beneath, a cased borehole. These examples demonstrated that the casing can improve detectability of a conductive target as compared to the scenario where no cased well is present.

With improvements in computing power, it has become possible to perform 3D numerical simulations with steel-cased wells. Simulations which capture the challenging geometry and large physical property contrasts due to well casings have have been successfully employed for DC and EM problems (e.g. \cite{Swidinsky2013, Commer2015, Hoversten2015, Tang2015, Um2015, Weiss2016, Yang2016, Heagy2018a}). These advances provide the opportunity to delve further into aspects of the physics governing the behavior of fields, fluxes, and charges when casings are present in an electrical or electromagnetic survey. To develop our understanding we start with DC resistivity.

In this paper, we focus our attention on three aspects of DC resistivity in the presence of steel-cased well. In Section \ref{sec:casing_integrity}, we examine the feasibility of conducting a surface DC survey to detect a flaw in the casing and discuss factors that influence detectability of a flaw. In Section \ref{sec:survey_design}, we examine the use of DC resistivity for geophysical imaging when a steel-cased well is present. Finally, in Section \ref{sec:approximating_wells}, we assess strategies applied in the literature for approximating a steel-cased well with a coarse-scale model to reduce computational cost. We focus our efforts on examining the finer details of the physics and hence we will consider only models with a single well in our simulations. We refer readers to \cite{Weiss2017} for discussion of DC resistivity simulations with multiple wells.

Source codes for all of the simulations shown are open source, licensed under the MIT license, and are available as Jupyter notebooks at: https://github.com/simpeg-research/heagy-2018-dc-casing \citep{Heagy2018b}. The examples in the paper have been selected with an emphasis on examining physical principles; however, we envision that the Jupyter notebooks included with this publication could serve as useful survey design tools.

\section{Governing equations and numerical modelling}

The governing equations for the DC resistivity problem are given by:
\begin{equation}
\begin{split}
\nabla \cdot \vec{j} &= I\left(\delta(\vec{r} - \vec{r}_{s^{+}}) - \delta(\vec{r} - \vec{r}_{s^{-}})\right) \\
\vec{e} &= - \nabla \phi
\end{split}
\label{eq:dc_equations}
\end{equation}
where $\vec{j}$ is the current density, $I$ is the magnitude of the source current, and $\vec{r}_{s^+}$ and $\vec{r}_{s^-}$ are the locations of the positive and negative source electrodes, respectively. In the electrostatic limit, which is applicable for the DC experiment, the electric field $\vec{e}$ is curl-free and can therefore be expressed as the gradient of a scalar potential $\phi$, giving the second equation in equation \ref{eq:dc_equations}. The electric field and the current density are related through Ohm’s law:
\begin{equation}
\vec{j} = \sigma \vec{e}
\label{eq:ohms_law}
\end{equation}
which we invoke to reduce the two first-order partial differential equations in equation \ref{eq:dc_equations} to a single, second order equation in $\phi$:
\begin{equation}
\nabla \cdot \sigma \nabla \phi = - I\left(\delta(\vec{r} - \vec{r}_{s^{+}}) - \delta(\vec{r} - \vec{r}_{s^{-}})\right)
\label{eq:dc_equation_second_order}
\end{equation}
In addition to considering the current density and electric fields, we will also present results in terms of charges. The charge density is related to the electric field through
\begin{equation}
\nabla \cdot \vec{e} = \frac{\rho_f}{\varepsilon_0}
\label{eq:charge_density}
\end{equation}

\begin{figure}
    \begin{center}
    \includegraphics[width=0.6\textwidth]{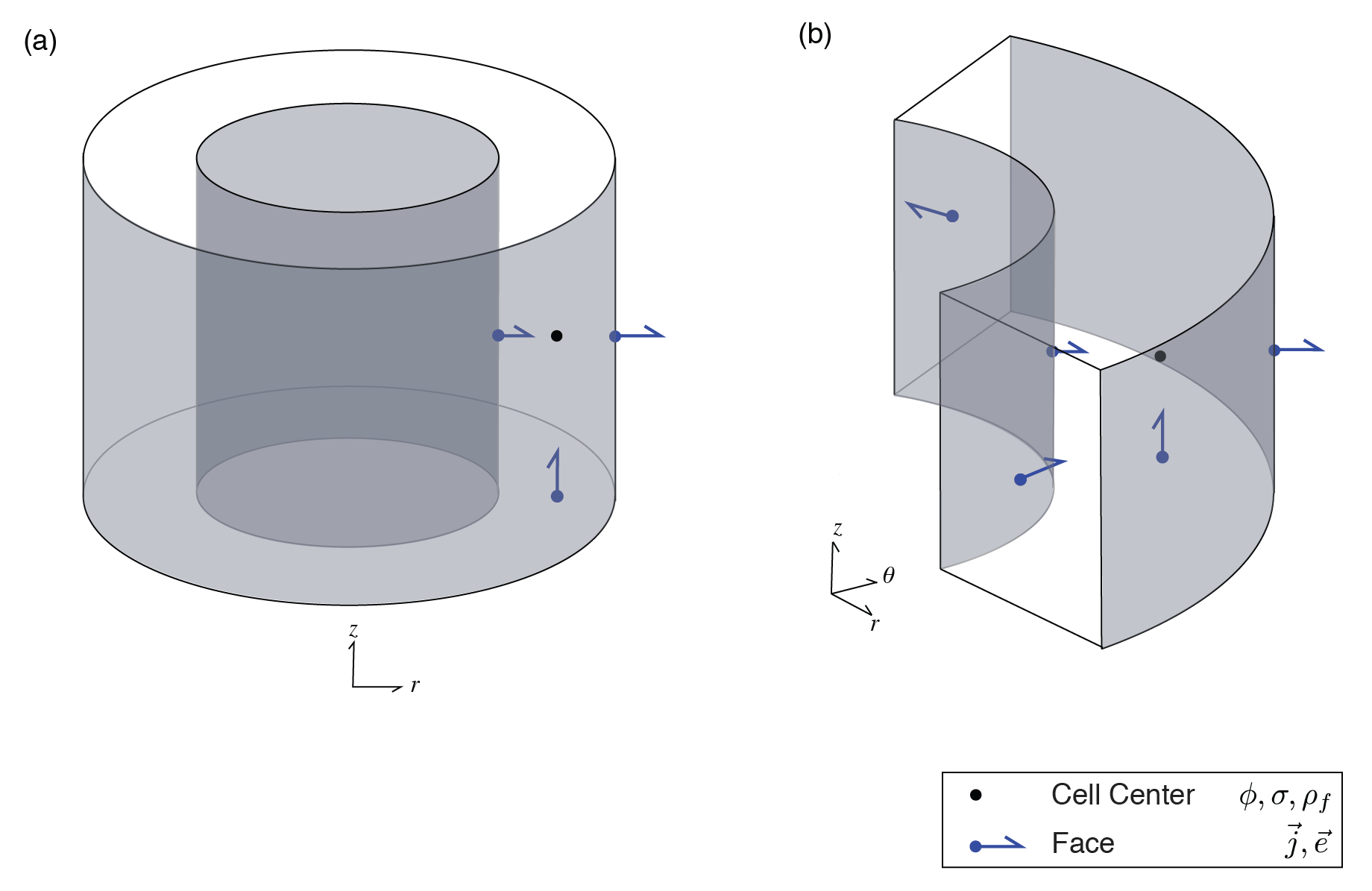}
    \end{center}
\caption{
    Cylindrical finite volume cells for a cell-centered discretization of the DC resistivity problem
    : (a) a cylindrically symmetric cell , (b) a 3D cylindrical cell.
    Scalar quantities ($\phi$, $\sigma$, $\rho_f$) are discretized at cell centers and vector
    quantities ($\vec{j}$, $\vec{e}$) are computed on cell faces.
}
\label{fig:dc_discretization}
\end{figure}

To numerically solve equation \ref{eq:dc_equation_second_order}, we use a finite volume approach, with the electric potential and the electrical conductivity discretized at cell centers. From the discrete solution for the electric potentials, the discrete electric field, current density and charge density can be computed directly. The vector quantities (electric field and current density) are computed on cell faces while the charge density is computed at cell centers. We employ both cylindrically symmetric and 3D cylindrical meshes, which include an azimuthal discretization. Figure \ref{fig:dc_discretization} demonstrates the discretization on:  (a) a cylindrically symmetric mesh and (b) a 3D cylindrical mesh.

All of the numerical simulations are run with the open source software described in \cite{Heagy2018a}, which relies on the electromagnetics module within SimPEG \citep{Cockett2015, Heagy2017}. In \cite{Heagy2018a}, we demonstrate validation of the code by comparing a time-domain EM simulation, which uses the same DC resistivity forward simulation code as used in this paper to compute the initial condition, with solutions presented in \cite{Commer2015} as well as with the Finite Volume OcTree code described in \cite{Haber2007}.

\section{DC resistivity for casing integrity}
\label{sec:casing_integrity}

Degraded or impaired wells can pose environmental and public-health hazards. A flaw in the cement or casing can provide a conduit for methane to migrate from depth into groundwater aquifers or into the atmosphere. This is particularly of concern for shale gas wells. Elevated levels of thermogenic methane, which are attributed to deep sources (rather than biogenic methane which can be generated closer to the surface), in groundwater wells in Pennsylvania has been positively correlated with proximity to shale gas wells in the Marcellus and Utica \citep{Osborn2011, Jackson2013}, and failure rates of unconventional wells (e.g. shale gas wells) are estimated to be 1.57 times larger than those of conventional wells drilled in the same time-period \citep{Ingraffea2014}. Wells can fail if there is a compromise in the cement or the casing. To diagnose the integrity of a well with electrical methods, we require a contrast in electrical conductivity to be associated with the flaw, thus we will focus our attention to detecting flaws in the highly conductive casing.

Under what circumstances should we be able to detect a flaw in the casing using DC resistivity from the surface? To address this question, we begin by examining how a flaw which comprises the entire circumference of the pipe along some depth interval changes the charge distribution and thus the resultant electric fields we measure on the surface. From there, we investigate the role of parameters including the depth of the flaw and the background conductivity on our ability to detect it from the surface. Finally, we examine the scenario in which only a portion of the circumference of the pipe is flawed.

\subsection{Basic experiment}

The experiment we consider is a ``top-casing'' DC resistivity experiment where one electrode is connected to the wellbore at the surface and a return electrode is positioned some distance away. The concept and basic physics is the same as a mise-\`a-la-masse survey in which the positive electrode is connected to a conductive target. When the source is turned on, positive charges are distributed on the interface between the conductive target and the resistive host. Electric potentials are measured on the surface and these data are then used to infer information about the extent of the conductor \citep{Telford1990}. Applying the same principles to a casing integrity experiment, we connect a positive electrode to the casing, and for an intact casing, positive charges will be distributed on the outer interface of the casing along its entire length. If corrosion causes a flaw across the diameter of the casing, the continuity of the conductive flow-path for charges is interrupted. Thus, we expect a larger charge to reside on the top portion of the flawed casing than would be observed if the casing were intact. As a result, the electric field observed at the surface should be larger than if the casing were intact. The difference in electric field (or electric potentials) from the expected electric field that results from an intact well could then be an indicator that there is a problem with the well.

\begin{figure}
    \begin{center}
    \includegraphics[width=0.8\textwidth]{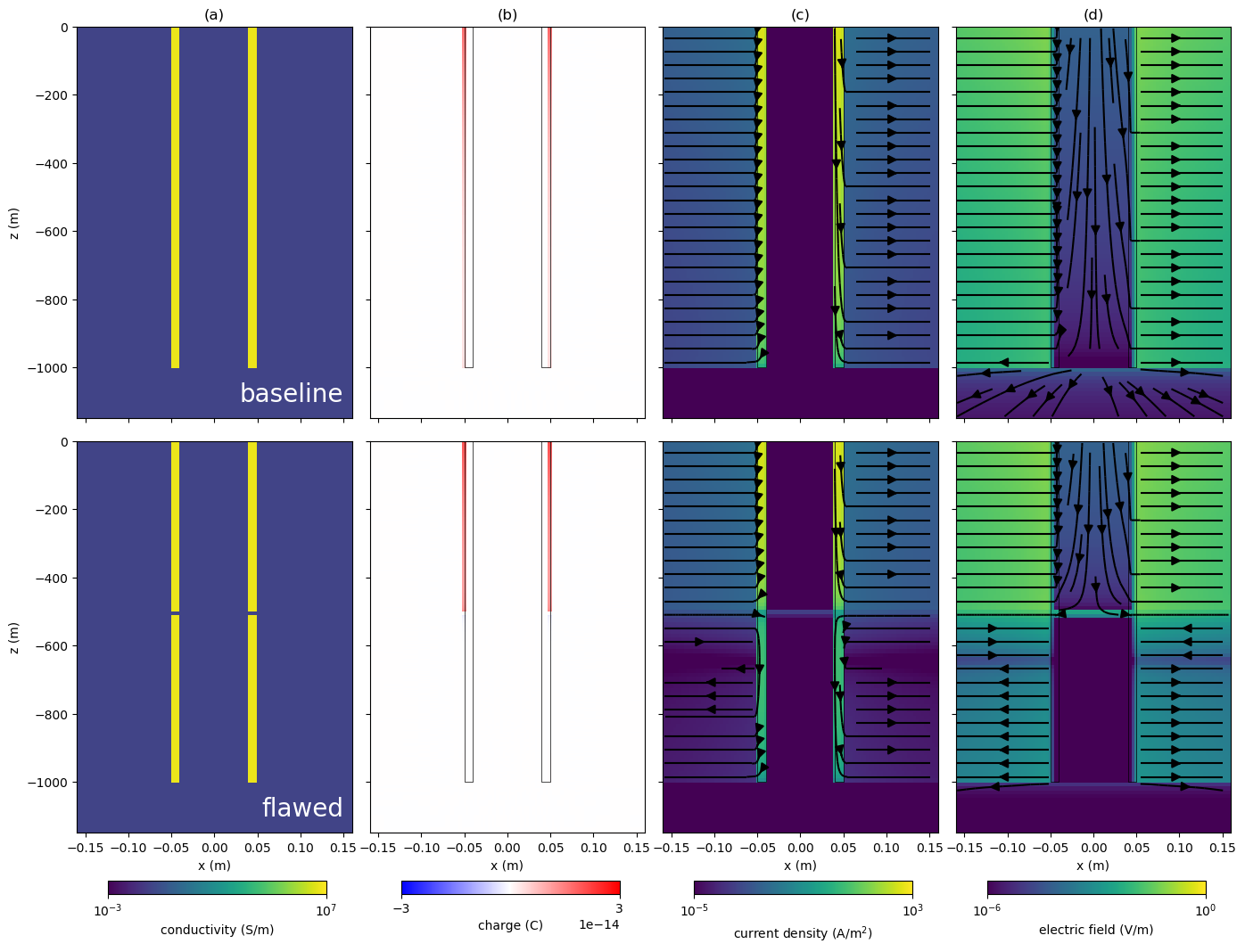}
    \end{center}
\caption{
    Cross section showing: (a) electrical conductivity, (b) current density, (c) charge density, and (d) electric field
    for a top-casing DC resistivity experiment over (top) an intact 1000m long well and (bottom) a 1000m long well
    with a 10m flaw at 500m depth.
}
\label{fig:casing_integrity_basics}
\end{figure}

To demonstrate the principles, we consider a simple model of a casing in a half-space. The intact well is 1 km long, has an outer diameter of 10 cm, a thickness of 1 cm and a conductivity of $5\times10^6$ S/m. The background is $10^{-1}$ S/m, and the conductivity of the inside of the well is taken to be equal to that of the background. The positive electrode is connected to the top of the casing and the return electrode is positioned 2 km away. To simulate the physics, the 3D cylindrical DC code described in \cite{Heagy2018a} was employed. In Figure \ref{fig:casing_integrity_basics} we show cross-sections of the: (a) electrical conductivity model, (b) current density, (c) charge density, and (d) electric field. The top row shows the intact well and the bottom row shows a flawed well which contains a 10 m gap in the casing at 500 m depth. As expected, the introduction of a resistive flaw prevents currents from reaching the bottom portion of the well. This results in increased currents, charge density, and thus electric fields within the top 500m.

To quantify the charge along the length of the well, we have plotted the charge as a function of depth for the intact well (black), flawed well (blue), and also a ``short'' well of 500 m length (grey dash-dot) in Figure \ref{fig:casing_charge}a. In each of the wells, we observe that there is an increase in charge density near the end of the discontinuity along the length of the well. This was also noted  in \cite{Griffiths1997} and \cite{Heagy2018a} and is attributed to edge-effects. At an interface between materials with two different conductivities, the normal component of the current density must be conserved, as well as the tangential component of the electric field; the discontinuity at the end of the pipe, and at the location of the flaw, means the continuity conditions must be preserved simultaneously in the radial and vertical directions, and this complicates the behavior of the fields, fluxes and charges. Another observation is that the flawed and short wells have nearly identical charge distributions in the top 500 m. In the bottom portion of the flawed well, where the remaining conductive material is, a small dipolar charge is introduced. However, this charge is nearly an order of magnitude smaller than the charge in the top portion of the pipe. This behavior was similarly noted by \cite{Wilt2018, Wilt2018a} in their examination of currents along the length of an intact and a flawed well.
The signal due to the flaw can be defined as the difference between the total response due to a flawed well and the total response due to an intact well (the primary); we will refer to this difference as the secondary response. The secondary charge is dipolar in nature with positive charge above the flaw and negative charge beneath the flaw. We note that the charge distributions along the short well, truncated where the flaw starts at 500 m depth, and along the top portion of the flawed well are almost identical; these charges are the source of signal for a surface electric field measurement. This suggests that an inversion strategy, where one attempts to estimate the length of a well, may be an effective approach for characterizing the depth to a flaw.

\subsubsection{Impact of the vertical extent of the flaw}

\begin{figure}
    \begin{center}
    \includegraphics[width=\textwidth]{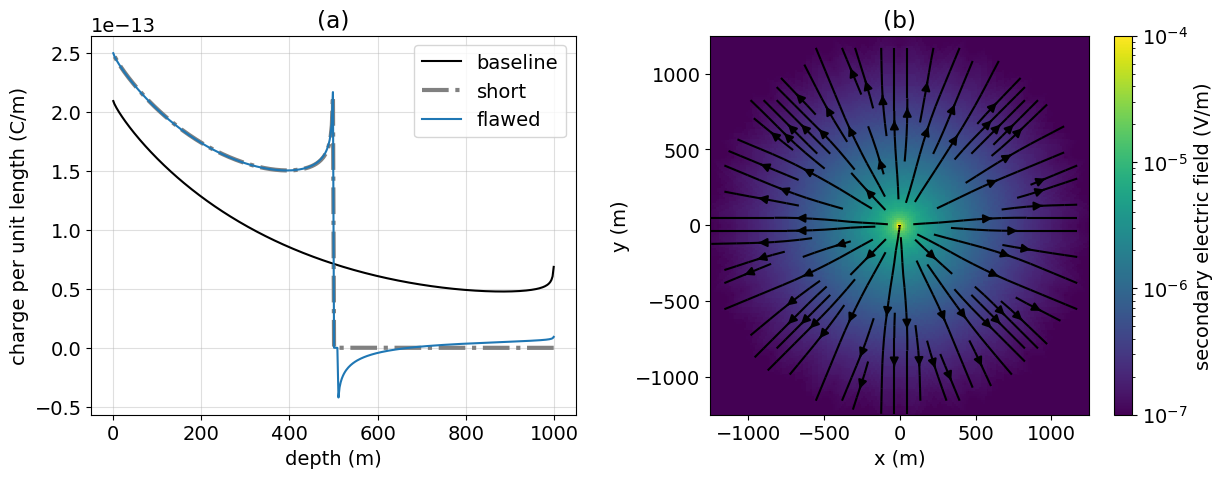}
    \end{center}
\caption{
    (a) Charge along the length of the intact well (black),
    a 500m well ( ``short'', grey dash-dot), and
    a well with a 10m flaw at 500m depth (blue),
    in a top-casing DC resistivity experiment.
    (b) Secondary electric field due on the surface of the earth due to the
    flaw in the casing. The primary is
    defined as the electric field due to the 1000m long intact well. The return electrode
    is 2000m away from the well.
}
\label{fig:casing_charge}
\end{figure}

A 10 m flaw is quite long and it is of interest to see how the results are changed if the flaw has a smaller vertical extent. The distribution of charges shown in Figure \ref{fig:casing_charge} hints that the flaw may not need to be very long in order to still significantly influence the response. To confirm this, we adopt a much finer vertical discretization in order to model smaller flaws. Here, we use a shorter, 50 m long well in order to reduce computational load. The flaw is positioned at 25 m depth, and the length of the impairment is varied. This simulation is conducted on a cylindrically symmetric mesh. The positive electrode is connected to the casing, and a return electrode is positioned 50m away.

\begin{figure}
    \begin{center}
    \includegraphics[width=0.7\textwidth]{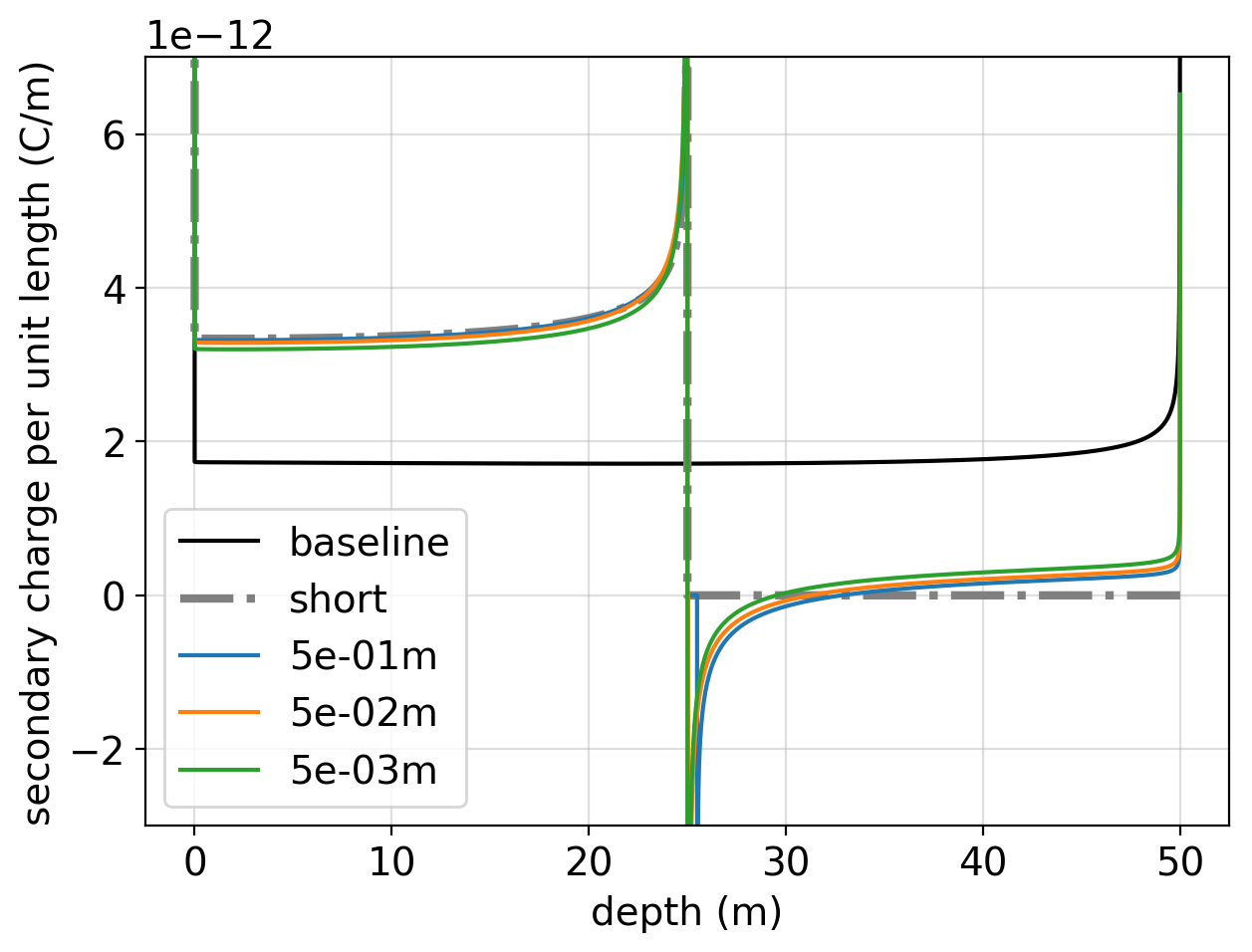}
    \end{center}
\caption{
    Charge along the length of a 50m long intact well (black),
    a 25m well (``short'', grey dash-dot), and four wells, each with a flaw
    starting at 25m depth and extending the length indicated by the legend
    ($5 \times 10^{-1}$ m (blue), $5 \times 10^{-2}$ m (orange), and $5 \times 10^{-3}$ m (green))
    in a top-casing DC resistivity experiment.
    For reference, the diameter of the casing is $10^{-1}$ m and its thickness is $10^{-2}$ m.
    The return electrode
    is 50m away from the well and a cylindrically symmetric mesh was used in the simulation.
}
\label{fig:casing_charge_flawdz}
\end{figure}

The resultant charge distributions are shown in Figure \ref{fig:casing_charge_flawdz}. For comparison, we have again shown the charge on a well that is truncated at the location of the flaw (grey dash-dot line). The charge distribution is similar for all of the flawed-well scenarios, even for flaws smaller than the thickness of the casing ($10^{-2}$ m). We see similar behavior to that shown in Figure \ref{fig:casing_charge}, where positive charge accumulates within the top portion of the well and a small dipole charge is present in the bottom portion of the well. There are minor differences in amplitude as the vertical extent of the flaw is changed. As the extent of the flaw decreases, the amplitude of the dipolar charge on the bottom portion of the well increases slightly while the amplitude of the positive charge on the top portion of the well decreases. These distinctions, however, are small in magnitude, and even if the background is more conductive, the casing is still orders-of-magnitude larger in conductivity than any geologic material we are likely to encounter. Thus, we can conclude that, so long as the impairment affects the entire circumference of the casing, the extent of that flaw has little impact on the charge that accumulates in the top portion of the well. As such, we will proceed in our analysis using a 10 m flaw in the 1 km well so that a fine vertical discretization is not necessary.

\subsection{Survey design considerations}

When examining detectability of a signal, there are two aspects to consider: (1) the signal must be larger than the noise floor of the instrument, and (2) the signal must be a significant percentage of the primary; for the casing integrity experiment, the primary is the signal due to the intact well. Due to the cylindrical symmetry of the charge on the well, we expect the electric field at the surface to be purely radial, thus only radial electric field data need be collected at the surface.

\begin{figure}
    \begin{center}
    \includegraphics[width=\textwidth]{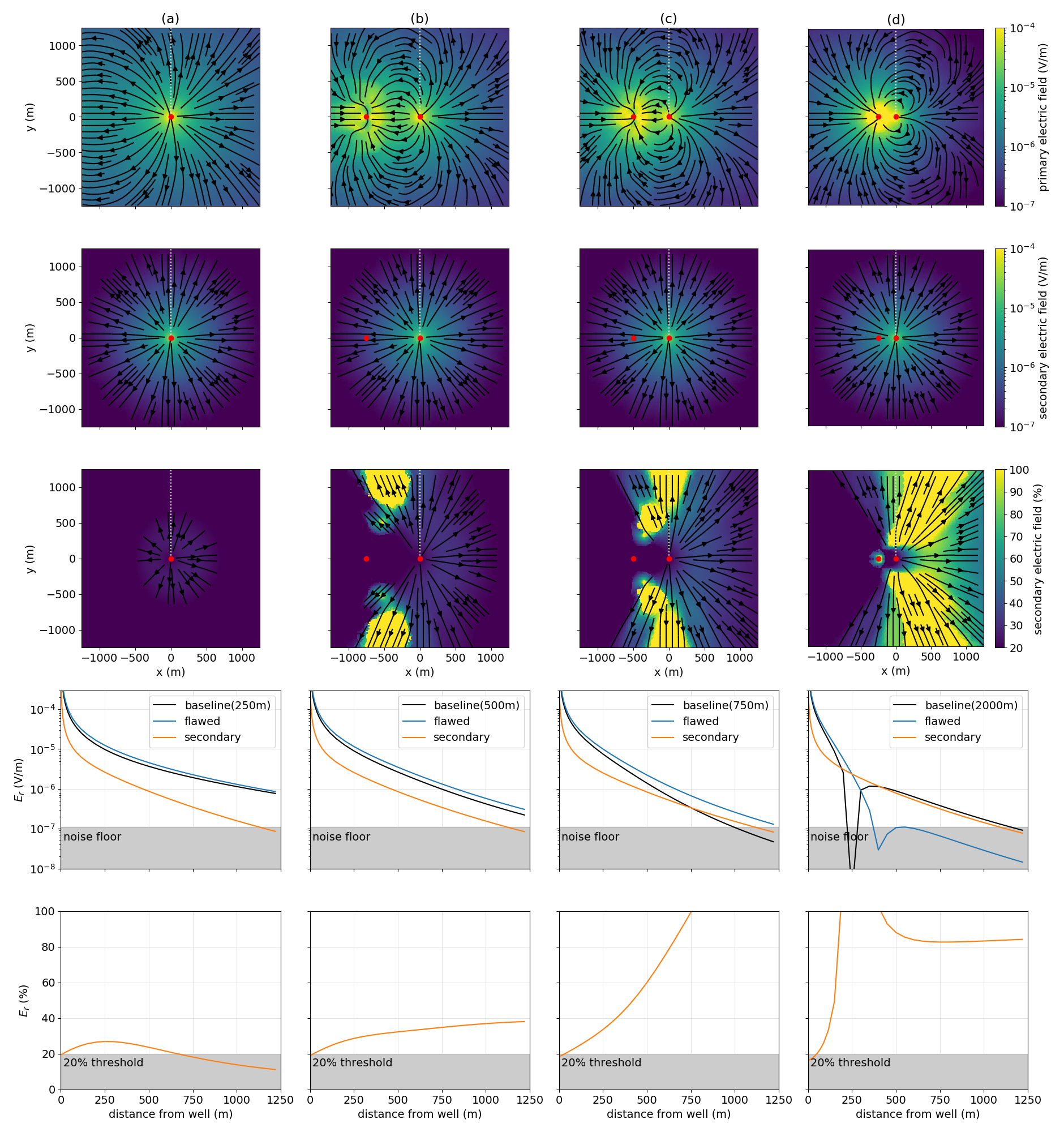}
    \end{center}
\caption{
    (Top row) primary electric field, (second row) secondary electric field,
    and (third row) secondary electric field as a percentage of the primary radial electric field
    for a return electrode that is offset (a) 2000m, (b) 750m, (c) 500m, and (d) 250m
    from the well. The primary is defined as the response due to the 1000m
    long, intact well. In each figure, the electrode locations are denoted by
    the red dots. In the third row, the colorbar has been limited
    between 20\% and 100\%. The fourth and fifth rows show radial electric field data
    collected along the $\theta=90^\circ$ azimuth (the white dotted lines in
    the top three rows). The fourth row shows the primary (black line), the total
    electric field due to the flawed well (blue line), and the secondary
    radial electric field (orange line). The fifth row shows the secondary as a
    percentage of the primary.
}
\label{fig:integrity_e_fields}
\end{figure}

In Figure \ref{fig:integrity_e_fields}, we have plotted the primary field (top row), secondary field (second row) and secondary field as a percentage of the primary (third row) for four different return electrode locations. In (a), the return electrode is 2000m offset from the well, in (b) the offset is 750m, in (c) the offset is 500m, and in (d) the offset is 250m. In addition to the plan view images, we have plotted the primary electric field (black),  total electric field for the flawed well (blue) and secondary radial electric field (orange) along the $\theta = 90^\circ$ azimuth in the fourth row of Figure \ref{fig:integrity_e_fields}. The fifth row shows the secondary as a percentage of the primary.

At the furthest offset (Figure \ref{fig:integrity_e_fields}a), there is nearly complete cylindrical symmetry in the primary field. With complete cylindrical symmetry there is no preferential direction along which to collect data. As we move the return electrode closer, for example to 750 m from the well, we notice that the secondary electric field does not change substantially. However, if we examine the ratio of the secondary to the primary (second and fifth rows), we see that the ratio has increased. Although the primary field has similar, if not larger, amplitude near the well, it also has considerable curvature. As a result, the proportion of the primary field that is in the radial direction has decreased in amplitude. Hence the important characteristic, the ratio of the secondary to primary of the radial components, has increased. The above principles are further enhanced as the return current is brought closer to the well as in panels (c) and (d), where the return electrode is brought to 500 m and 250 m from the well.  Again, for all of these examples the amplitude of the secondary field at the surface is quite similar. However, the choice of azimuth for the survey line will greatly affect the size of the ratio. In terms of survey design, we can take advantage of the return electrode to reduce coupling with the primary.

For the examples that follow, we will place the return electrode at 500 m from the well and collect radial data along a line that is perpendicular to the source-line. We will examine several factors influencing detectability of a flaw, including the depth of the flaw and the conductivity of the background in the following sections. We will also examine the scenario where only a portion of the circumference of the well has been compromised.

\subsection{Factors influencing detectability}
\subsubsection{Depth of the flaw}
\begin{figure}
    \begin{center}
    \includegraphics[width=0.8\textwidth]{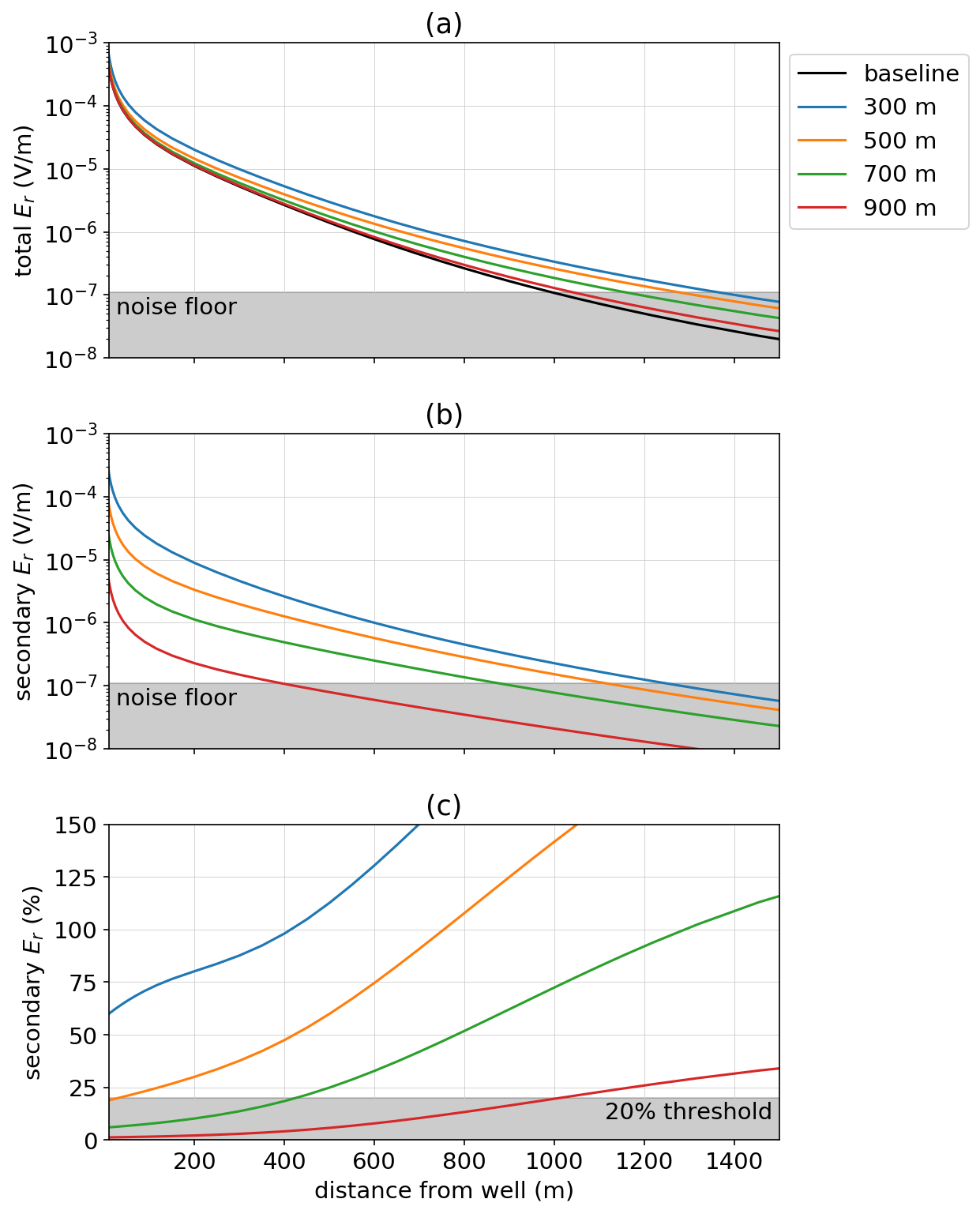}
    \end{center}
\caption{
    Radial electric field as the depth of the flaw along a 1km long well is varied.
    The positive electrode is connected to the top of the casing, the negative electrode
    is positioned 500m away and data are measured along a line $90^\circ$ from the
    source electrodes. In (a), we show the total electric field for four flawed wells,
    each with a 10m flaw at the depth indicated on the legend. The black line shows
    the radial electric field due to an intact well; we define this as the primary.
    In (b), the secondary radial electric field is plotted and in (c), we show the
    secondary radial electric field as a percentage of the primary.
}
\label{fig:integrity_depth}
\end{figure}

The introduction of a flaw in the well changes the distribution of charges along the length of the well and causes a secondary dipolar charge centered about the flaw. The position and strength of this dipole will affect our ability to detect the flaw. To examine this, we again use a model of a 1 km pipe in a $10^{-1}$ S/m background. The positive electrode is connected to the top of the well and a return electrode is 500 m away from the well. We vary the depth of a 10 m flaw from 300 m to 900 m. In Figure \ref{fig:integrity_depth}, we have plotted radial electric field results along a line perpendicular to the source electrodes. In (a), we show total radial electric field, in (b) the secondary radial electric field (with the primary being the electric field resulting from the intact well, shown in black in panel a), and in (c) we show the secondary radial electric field as a percentage of the primary. We have indicated where values fall below a $10^{-7}$ V/m noise floor on Figure \ref{fig:integrity_depth} (a) and (b), as well as those that fall below a 20\% threshold in (c). A threshold of 20\% may be conservative, however, it does depend on knowledge of the background conductivity as well as the geometry and physical properties of the well. In many scenarios, these may not be well-constrained, thus we select a conservative threshold for this analysis. Any detectability analysis will be site-dependent and we have therefore made all source-code available so that a similar workflow may be followed and adapted to include setting-specific parameters.

When a well is impaired, the total radial electric field is larger than that due to the baseline, intact well. The strength of the secondary response decreases as the depth of the flaw increases. For this example of a 1 km long well in a $10^{-2}$ S/m background, a flaw at 900 m depth is not detectable; there is no overlap between the region in which the secondary electric field (Figure \ref{fig:integrity_depth}b) is above the noise floor and the region in which the secondary comprises a significant percentage of the primary (Figure \ref{fig:integrity_depth}c). This might be expected, as the difference between the charges distributed along a 900 m long segment versus the 1 km long well are not drastically different. For a flaw at 700 m depth, there is a window between 400 m offset and 800 m offset over which the radial electric field data are sensitive to the flaw. As the depth to the impairment decreases, both the spatial extent over which data are sensitive to the flaw, and the magnitude of the secondary response in those data, increase.

\subsubsection{Background conductivity}

\begin{figure}
    \begin{center}
    \includegraphics[width=0.8\textwidth]{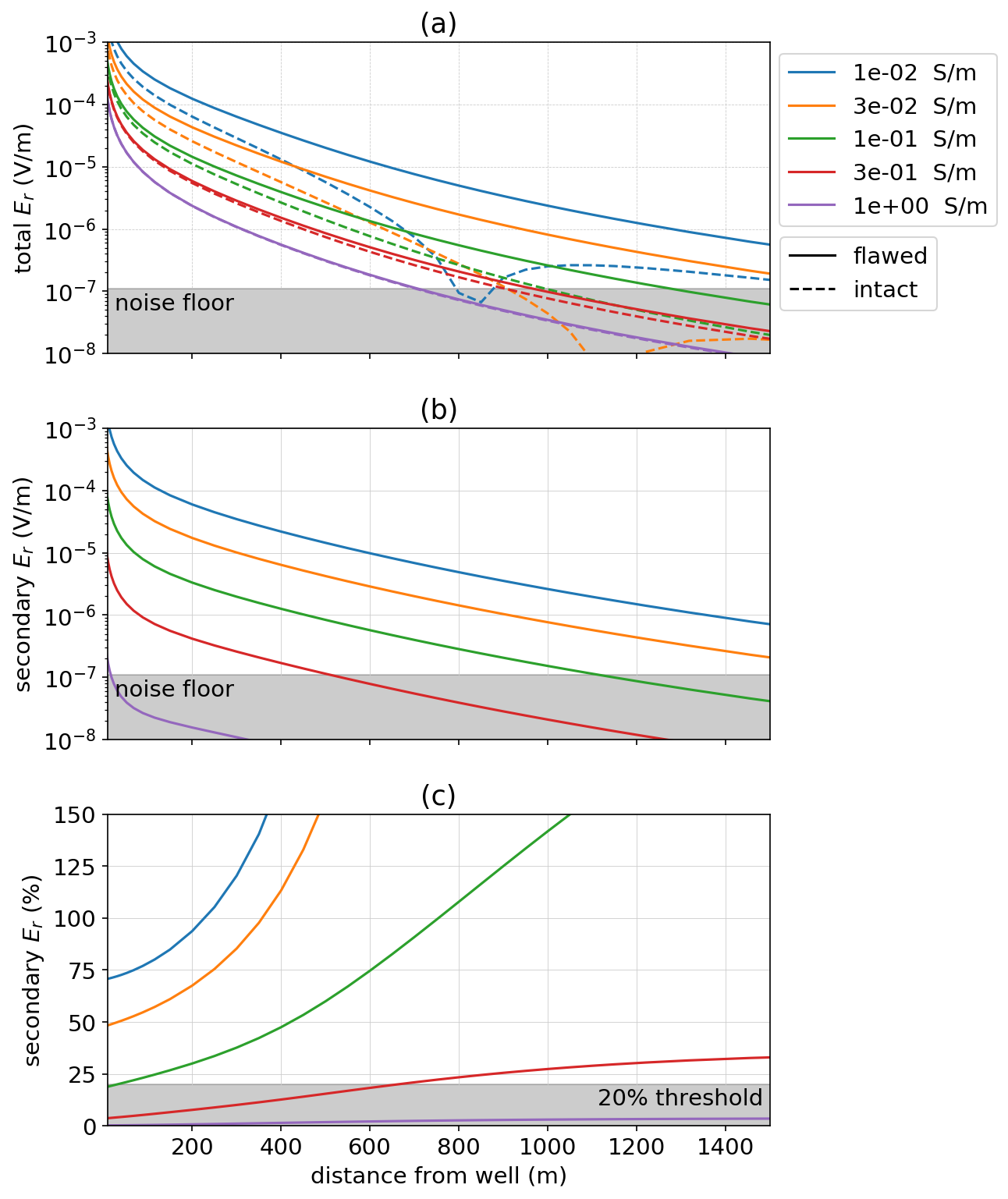}
    \end{center}
\caption{
    Radial electric field as the conductivity of the background is varied for a 1km well with a 10m flaw at 500m depth.
    The positive electrode is connected to the top of the casing, the negative electrode
    is positioned 500m away and data are measured along a line $90^\circ$ from the
    source electrodes. In (a), we show the total electric field for five different background conductivities,
    each indicated on the legend. The solid lines indicate the response of the flawed well and the dashed lines indicate the response of the intact well (the primary).
    In (b), the secondary radial electric field is plotted and in (c), we show the
    secondary radial electric field as a percentage of the primary.
}
\label{fig:integrity_conductivity}
\end{figure}

The total charge on the well is controlled by the contrast in conductivity between the steel-cased well and the surrounding geology. Increasing the conductivity of the background reduces that contrast thus reducing the amount of charge on the well. The result is a decrease in the total electric field at the surface. Similarly, the strength of the secondary dipolar charge introduced with the presence of an impairment also depends upon the available charge and will also be reduced with increasing background conductivity. In Figure \ref{fig:integrity_conductivity}, we have adopted the same model of a 1 km well with a 10 m impairment at 500 m depth, and show the radial electric field for the flawed (solid lines) and intact (dashed lines) wells as the background conductivity is varied. A resistive background promotes the strongest total and secondary signals. As the conductivity increases, detectability becomes more challenging; at a conductivity of $3 \times 10^{-1}$ S/m, the flaw at 500 m depth is undetectable as there is no overlap in the regions where the secondary signal is above the noise floor and where it comprises a significant percentage of the primary.

\begin{figure}
    \begin{center}
    \includegraphics[width=0.8\textwidth]{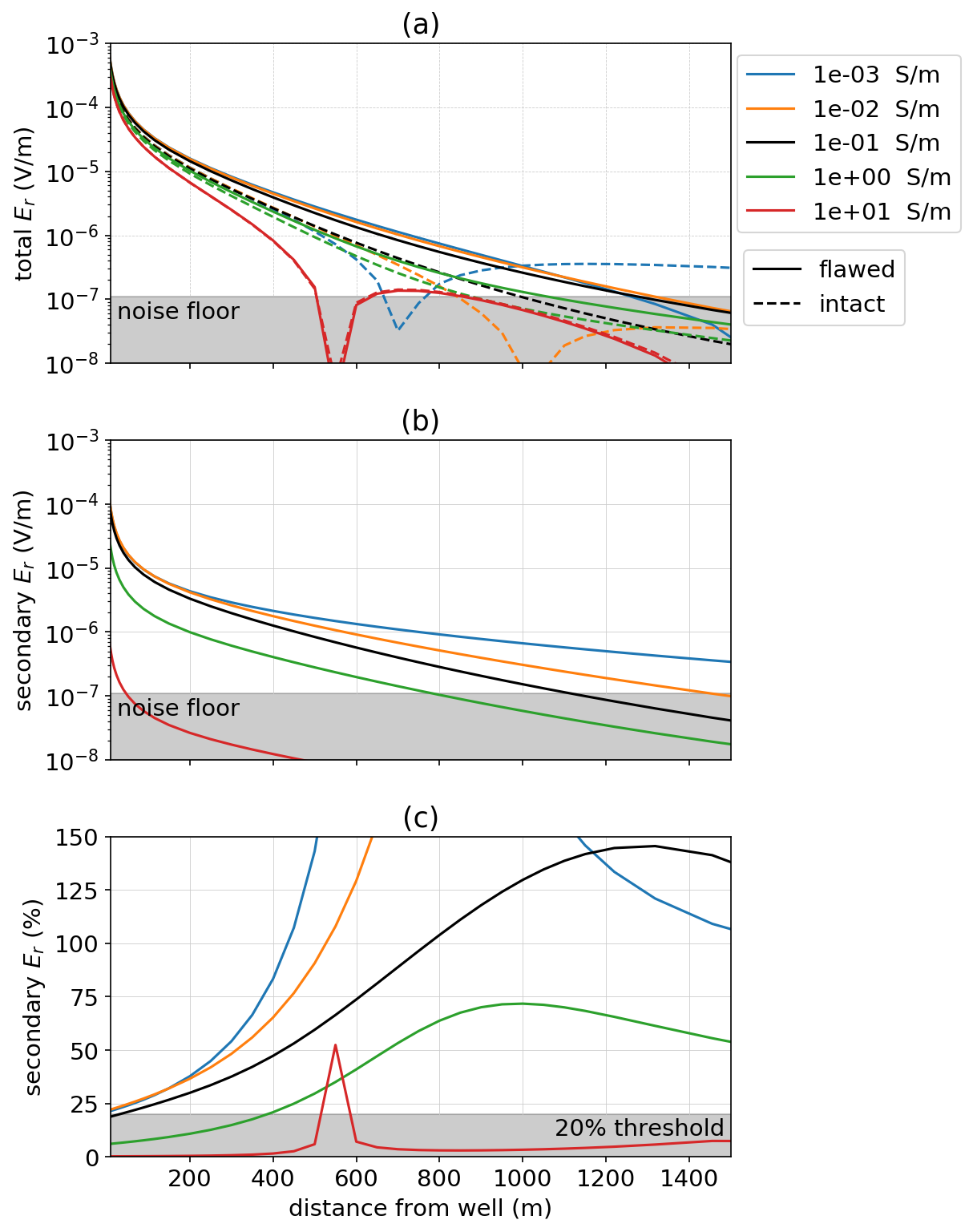}
    \end{center}
\caption{
    Radial electric field as the conductivity of a 50m thick layer positioned at 400m depth is varied.
    The positive electrode is connected to the top of the casing, the negative electrode
    is positioned 500m away and data are measured along a line $90^\circ$ from the
    source electrodes. In (a), we show the total electric field for five different layer conductivities.
    The black line shows the scenario where the layer has the same conductivity as the background.
    The dashed-lines indicate the intact well and the solid lines indicate the flawed well.
    In (b), the secondary radial electric field is plotted (with respect to an intact well primary)
    and in (c), we show the
    secondary radial electric field as a percentage of the primary.
}
\label{fig:integrity_layer}
\end{figure}

Variations in the background geology will also influence the distribution of charges and thus the measured signal at the surface. To examine the challenges introduced when variable geology is considered, we introduce a layer into the model and vary its conductivity. The layer is 50 m thick and its top is at 400 m depth. The flaw will again be positioned at 500 m depth, and the background conductivity is $10^{-1}$ S/m. The return electrode is 500 m from the well, and radial electric field data are measured along a line perpendicular to the source. In Figure \ref{fig:integrity_layer}, we show data for a flawed well (solid) and intact well (dashed) for scenarios in which a conductive or resistive layer is positioned above the flaw. The presence of a resistive layer improves detectability, while a conductive layer reduces detectability.

\begin{figure}
    \begin{center}
    \includegraphics[width=0.8\textwidth]{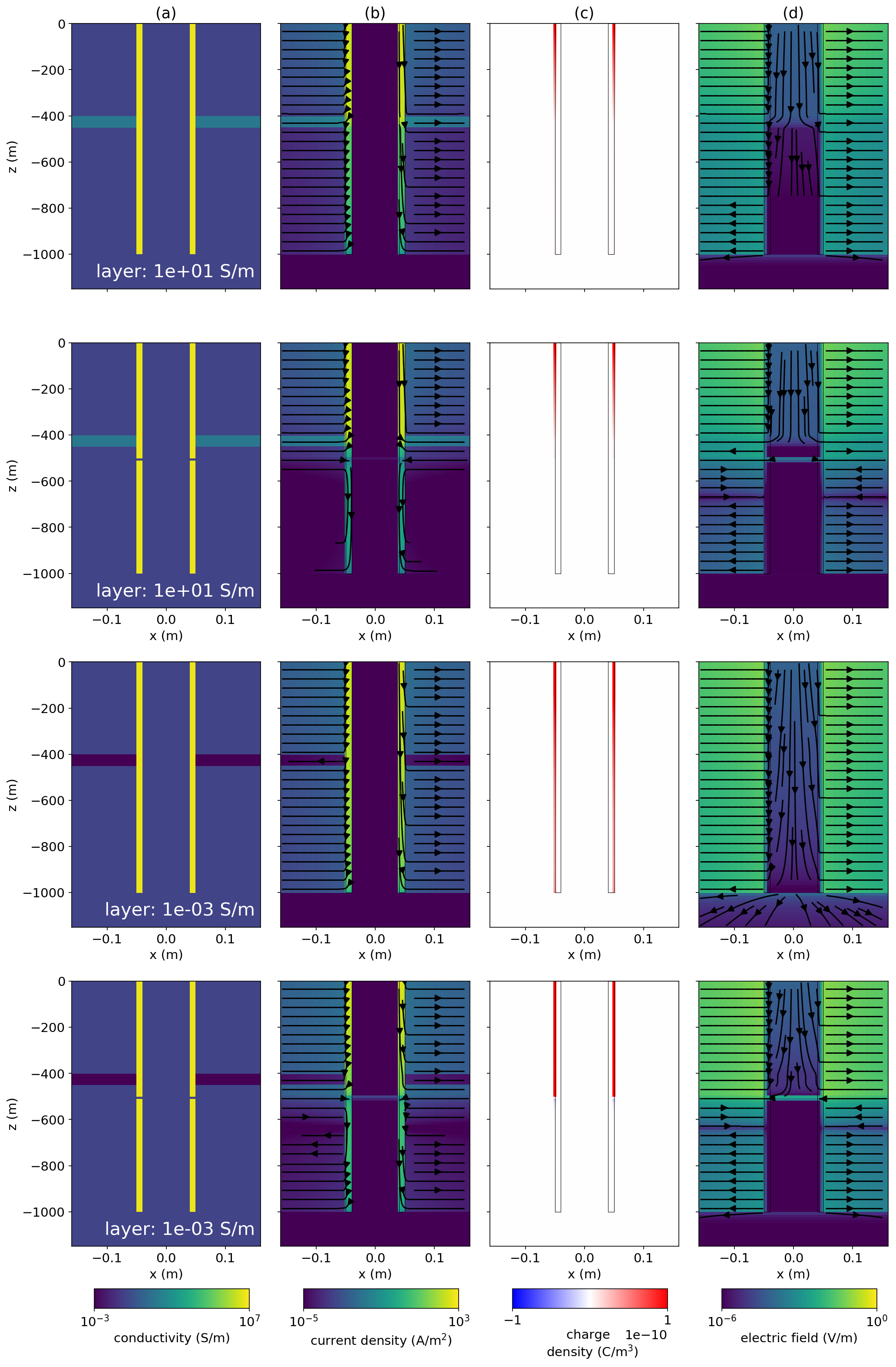}
    \end{center}
\caption{
    Cross section showing: (a) electrical conductivity, (b) current density, (c) charge density,
    and (d) electric field for a top-casing DC resistivity experiment over models with a conductive layer
    (top two rows) and a model with a resistive layer (bottom two rows). In all, the layer extends from
    400m to 450m depth. The plots in the second and fourth rows show the model, currents, charges and electric
    fields for a well with a 10m flaw at 500m depth.
}
\label{fig:integrity_layer_physics}
\end{figure}

To understand the physical phenomena governing this, we have plotted a cross section through: (a) the model, (b) the currents, (c) the charges, and (d) the electric field in Figure \ref{fig:integrity_layer_physics}. The first row shows the results for a model of an intact well with a conductive layer present and the second row shows the model with a flawed-well and a conductive layer. Similarly, the third and fourth rows show the results for an intact well and flawed well in a model with a resistive layer. In both examples, there is two orders of magnitude difference between the background and the layer. When a conductive layer is present, we see that it acts to ``short-circuit'' the system as there is significant current leak-off into that layer. This reduces the amount of current that reaches the flawed section of the well and decreases the total charge on the well, which is the source of our signal. Conversely, when a resistive layer is present, there is less leak-off of currents. In fact, \cite{Yang2016} showed that rather than leaking-off, currents can enter the casing if a resistive layer is present. In terms of detecting a flaw beneath a resistive layer, this means that the current density and charge along the well increases, thus amplifying the response due to the flaw.

\subsubsection{Conductivity of the casing}
\begin{figure}
    \begin{center}
    \includegraphics[width=\textwidth]{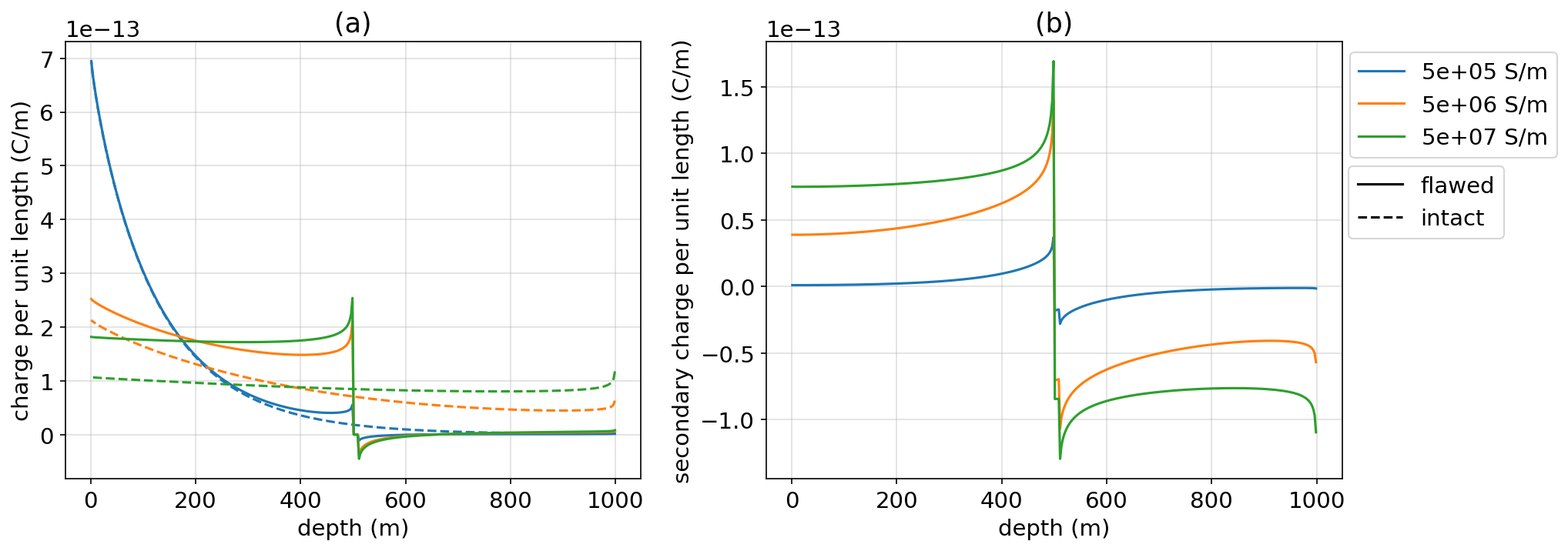}
    \end{center}
\caption{
    (a) Charge along the length of wells with three different
    conductivities (each indicated by a different color in the legend).
    The intact wells are denoted with dashed lines and the flawed wells
    are denoted with solid lines.
    (b) Secondary charge along the flawed and short wells. The primary is
    defined as the electric field due to the 1000m long intact well. The return electrode
    is 2000m away from the well.
}
\label{fig:casing_charge_sigma_casing}
\end{figure}

\begin{figure}
    \begin{center}
    \includegraphics[width=0.8\textwidth]{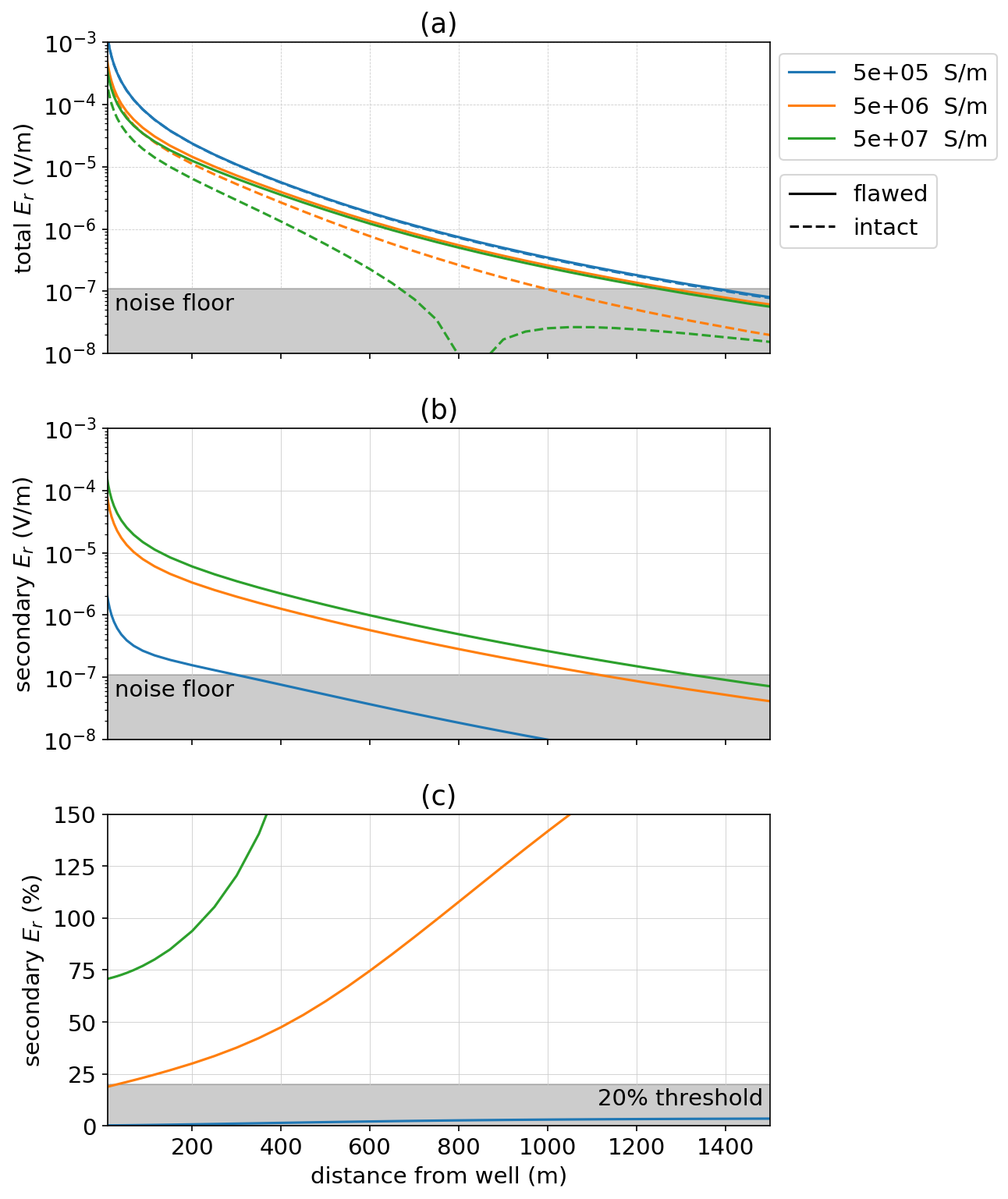}
    \end{center}
\caption{
    Radial electric field as the conductivity of the casing is varied for a 1km well with a 10m flaw at 500m depth.
    The positive electrode is connected to the top of the casing, the negative electrode
    is positioned 500m away and data are measured along a line $90^\circ$ from the
    source electrodes. In (a), we show the total electric field for three different casing conductivities,
    each indicated on the legend. The solid lines indicate the response of the flawed well and the dashed lines indicate the response of the intact well (the primary).
    In (b), the secondary radial electric field is plotted and in (c), we show the
    secondary radial electric field as a percentage of the primary.
}
\label{fig:integrity_conductivity_casing}
\end{figure}

The conductivity of the casing is also relevant to how the charges are distributed along its length. For highly conductive wells, the charge along the length of the well is approximately uniform. For more resistive wells, the charges follow an exponential decay, as shown in Figure \ref{fig:casing_charge_sigma_casing}. \cite{Schenkel1991} described the decay of currents, and thus the distribution of charges along the length of a well, in terms of the conduction length,
\begin{equation}
     \delta_L = \sqrt{\frac{S_c}{\sigma_0}} = \sqrt{\frac{2\pi r t \sigma_c}{\sigma_0}}
\label{eq:conduction_length}
\end{equation}
Where $S_c$ is the cross-sectional conductance of the casing ($S_c = 2\pi r t \sigma_c$ for a casing with radius $r$, thickness $t$, conductivity $\sigma_c$ and has units of [S $\cdot$ m]) and $\sigma_0$ is the conductivity of the background. The conduction length is akin to skin depth in electromagnetics and is the depth at which  the amplitude of currents have decreased by a factor of $e^{-1}$. Casing conductivities of $5 \times 10^5$ S/m, $5 \times 10^6$ S/m, and $5 \times 10^7$ S/m correspond to conduction lengths of $\sim 180$ m, $560$ m, $1800$ m. For the most resistive well shown, $5 \times 10^{5}$ S/m, the vast majority of current has decayed well before it reaches the flaw; the majority of charges are concentrated where the currents leak off, near the top of the well. Correspondingly, there is greater sensitivity to a flaw in a conductive well than in a resistive well, as is reflected in the radial electric field data shown in Figure \ref{fig:integrity_conductivity_casing}.

\subsubsection{Partial flaw}
The above examples considered an impairment that affects the entire circumference of the casing. This may be suitable in some scenarios where a particular geologic unit subjects the well to corrosive conditions, however, flaws may also be vertical cracks along the well (e.g. if pipe burst occurs). This is a much more challenging problem for DC resistivity because, if only a portion of the circumference is impaired, there is still a high-conductivity pathway for currents to flow along the entire length of the well. To examine the feasibility of detecting a partial flaw, we have run simulations where half of the circumference of the casing is compromised, leaving the other-half intact.

\begin{figure}
    \begin{center}
    \includegraphics[width=0.8\textwidth]{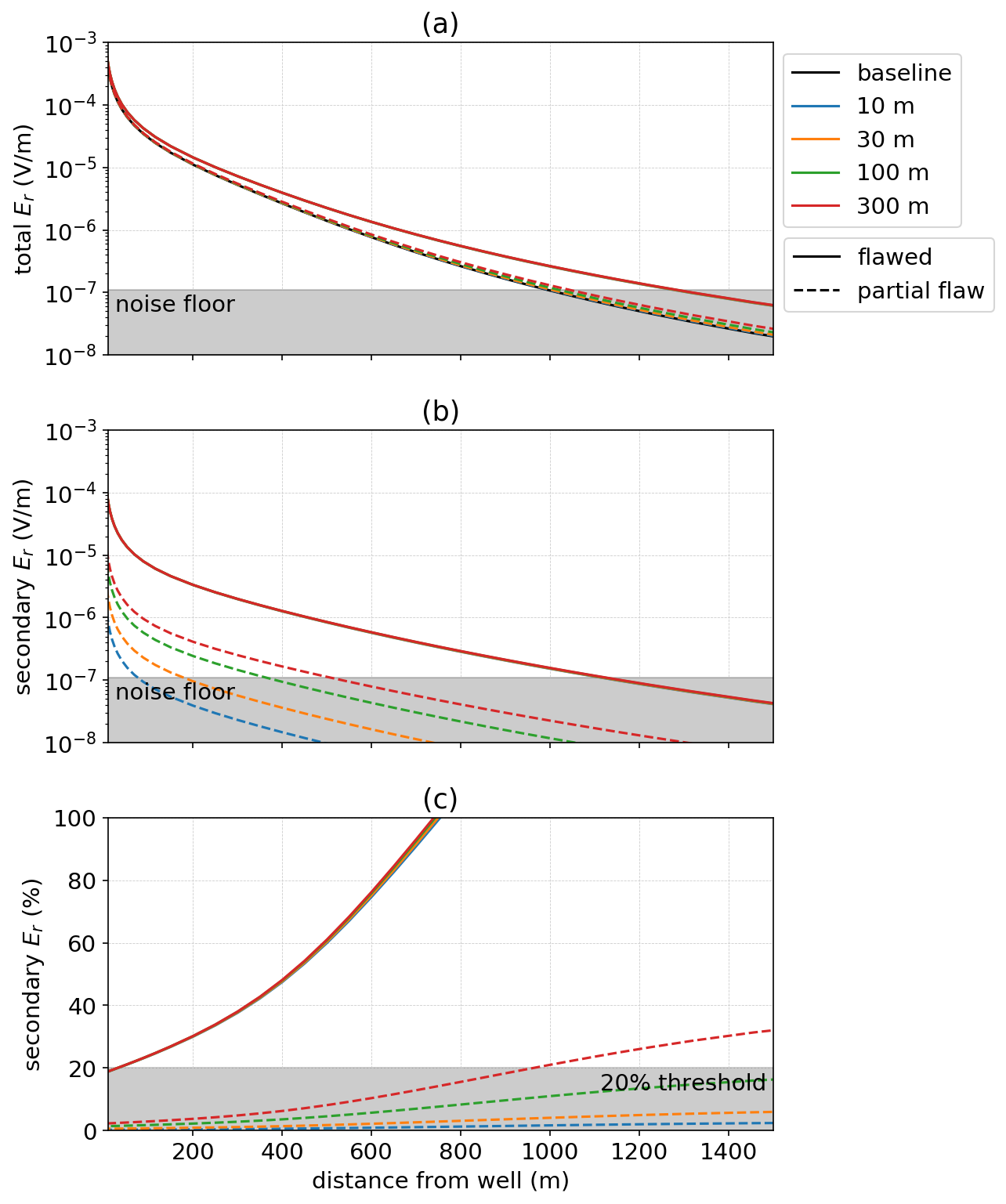}
    \end{center}
\caption{
    Radial electric field as the vertical extent of the flaw is varied.
    The positive electrode is connected to the top of the casing, the negative electrode
    is positioned 500m away and data are measured along a line $90^\circ$ from the
    source electrodes. In (a), we show the total electric field corresponding to four different flaw extents.
    The black line shows the response of the intact well.
    The dashed lines indicate the partially flawed wells (50\% of the circumference is compromised)
    and the solid lines flawed wells in which the entire circumference of the well has been compromised.
    In (b), the secondary radial electric field is plotted (with respect to an intact well primary)
    and in (c), we show the secondary radial electric field as a percentage of the primary.
}
\label{fig:integrity_partial_flaw}
\end{figure}

We consider four different depth extents of the flaw between 10m and 300m; in all scenarios the top of the flaw is at 500m. In Figure \ref{fig:integrity_partial_flaw}a, we have plotted the total radial electric field resulting from an intact well (black), wells where the entire circumference is compromised (solid) and wells in which 50\% of the circumference has been compromised (dashed).  Figure \ref{fig:integrity_partial_flaw} (b) and (c) show the secondary radial electric field and the secondary as a percentage of the primary, respectively.

These results show that the depth-extent of the flaw has little impact on the fully-compromised wells. This conclusion is consistent with the observations in our previous examples. However, if the well is partially flawed, we do see variation in the secondary response. By compromising 50\% of the circumference of the well, we have reduced the effective cross-sectional conductance over that portion of the well. Numerical experiments show that if instead of introducing a flaw which comprises 50\% of the circumference of the well, we reduce the conductivity of the intact well by 50\% over the same depth extent as the flaw, we obtain similar, but not identical, responses at the surface. Although for extensive flaws, there is a small region over which the secondary signal is above the noise floor, there are no regions where this coincides with measurements where the secondary fields are a significant percentage of the primary. There may be a subset of circumstances, such as if the flaw is near to the surface, or if the background geology is sufficiently well-known so that the percent threshold can be reduced, where a partial flaw may be diagnosed, however, these results demonstrate that a partial flaw is a challenging target for a DC resistivity survey.

In the next section, we transition from viewing the casing as the target to working on the scale of a geophysical imaging application in reservoir monitoring and viewing the casing as a high-conductivity feature present in that setting.

\section{Survey design for exciting targets at depth}
\label{sec:survey_design}
There are many problems in hydraulic fracturing, carbon capture and storage, and enhanced oil recovery that require targets to be illuminated and data to be acquired and inverted. Typically, these experiments include steel-cased wells and the target of interest could be resistive or conductive. The target could be immediately adjacent to a well or offset from it, and the survey may employ electrodes on the surface or positioned down-hole. Similarly, receivers may be positioned on the surface or in adjacent boreholes. Prior to designing a suitable inversion algorithm for imaging a target, we must first establish an understanding of how each of these factors influences our ability to detect a target in our data.

Detectability of a target requires two steps: (1) source fields must excite the target, and (2) receivers must be positioned so that the secondary response is measurable. In this section, we focus our attention on the first point, exciting the target. We will examine the impact of source electrode locations, the physical properties of the target and the geometry of the target on our ability to excite a response.
\subsection{Source location}
We begin by examining the impact of the source electrode location on our ability to deliver current to a region of interest in the model. We consider a 1km long well in a $10^{-1}$ S/m background. The well has a conductivity of $5 \times 10^6$ S/m, an outer diameter of 10cm thickness, and a 1cm thickness; these are the same parameters used for the casing integrity experiment described in the previous section. The conductivity of the fluid filling the casing is identical to that of the background. We are interested in effects near the well and thus the modeling can be carried out using the 2D cylindrical mesh provided that the return electrode is sufficiently far away. The return electrode is physically a disc of current at a radius equal to the distance of the return electrode from the well, in this case 2km. The assumption of cylindrical symmetry and the use of a distant return electrode has similarly been applied in \cite{Schenkel1991}.

\begin{figure}
    \begin{center}
    \includegraphics[width=\textwidth]{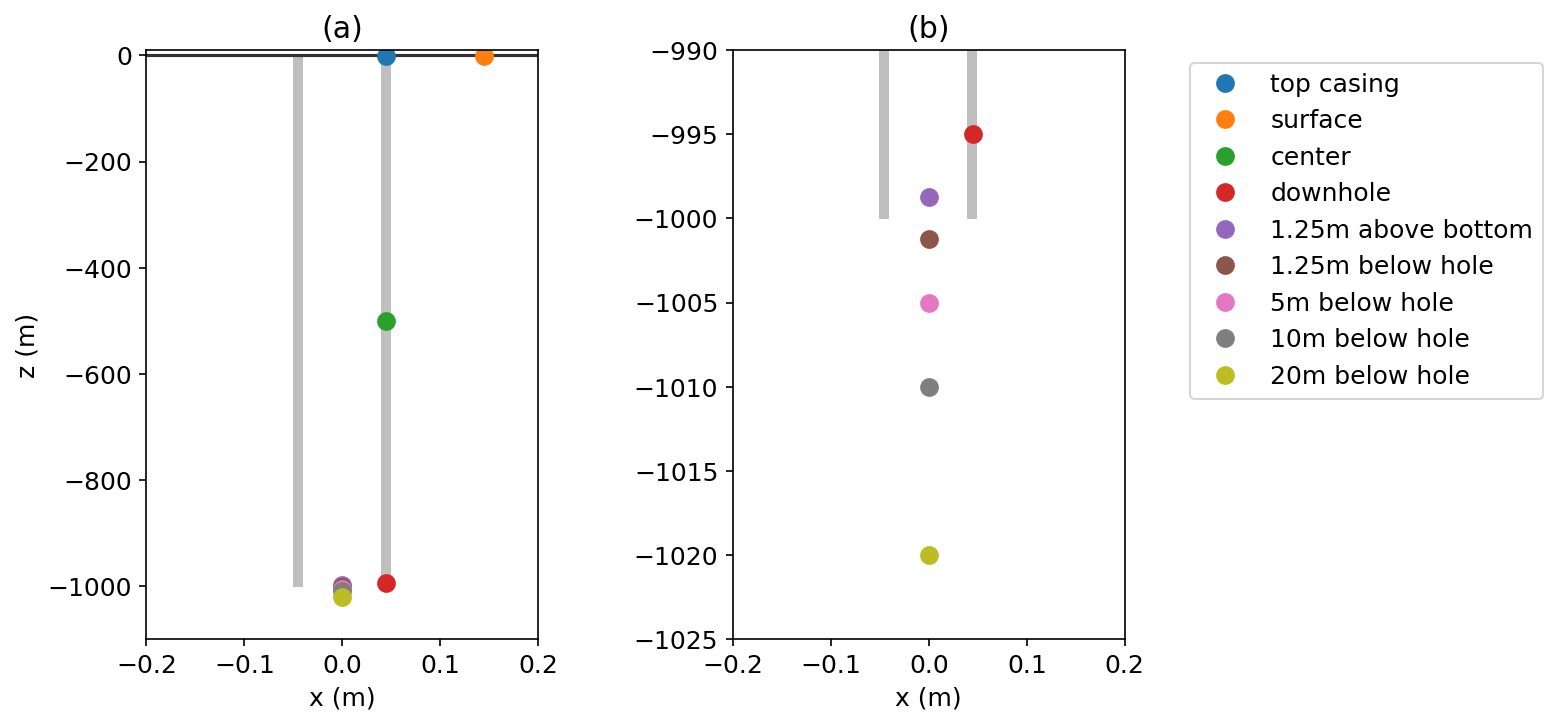}
    \end{center}
\caption{
    Electrode locations to be compared. The top casing electrode (blue),
    centered electrode (green, 500m depth), and downhole electrode (red, 500m depth)
    are connected to the casing. The surface electrode (orange) is offset from the well
    by 0.1m. The remaining electrodes are positioned along the axis of the casing. Panel (a)
    shows the entire length of the casing, while (b) zooms in to the bottom of the casing
    to show the separation between the electrodes beneath the casing.
}
\label{fig:electrode_location}
\end{figure}

To examine the impact that the source electrode location has on our ability to excite a target, we consider the five electrode locations shown in Figure \ref{fig:electrode_location}. Three of the electrodes are connected to the casing (tophole - blue, centered - green, and downhole - red); the remaining electrodes are  not connected to the casing; these include the surface electrode (orange) as well as the five electrodes near the end of the pipe (purple - within the pipe, brown, pink, grey and yellow are beneath the end of the pipe). The surface electrode is offset from the well by 0.1m.

\begin{figure}
    \begin{center}
    \includegraphics[width=0.8\textwidth]{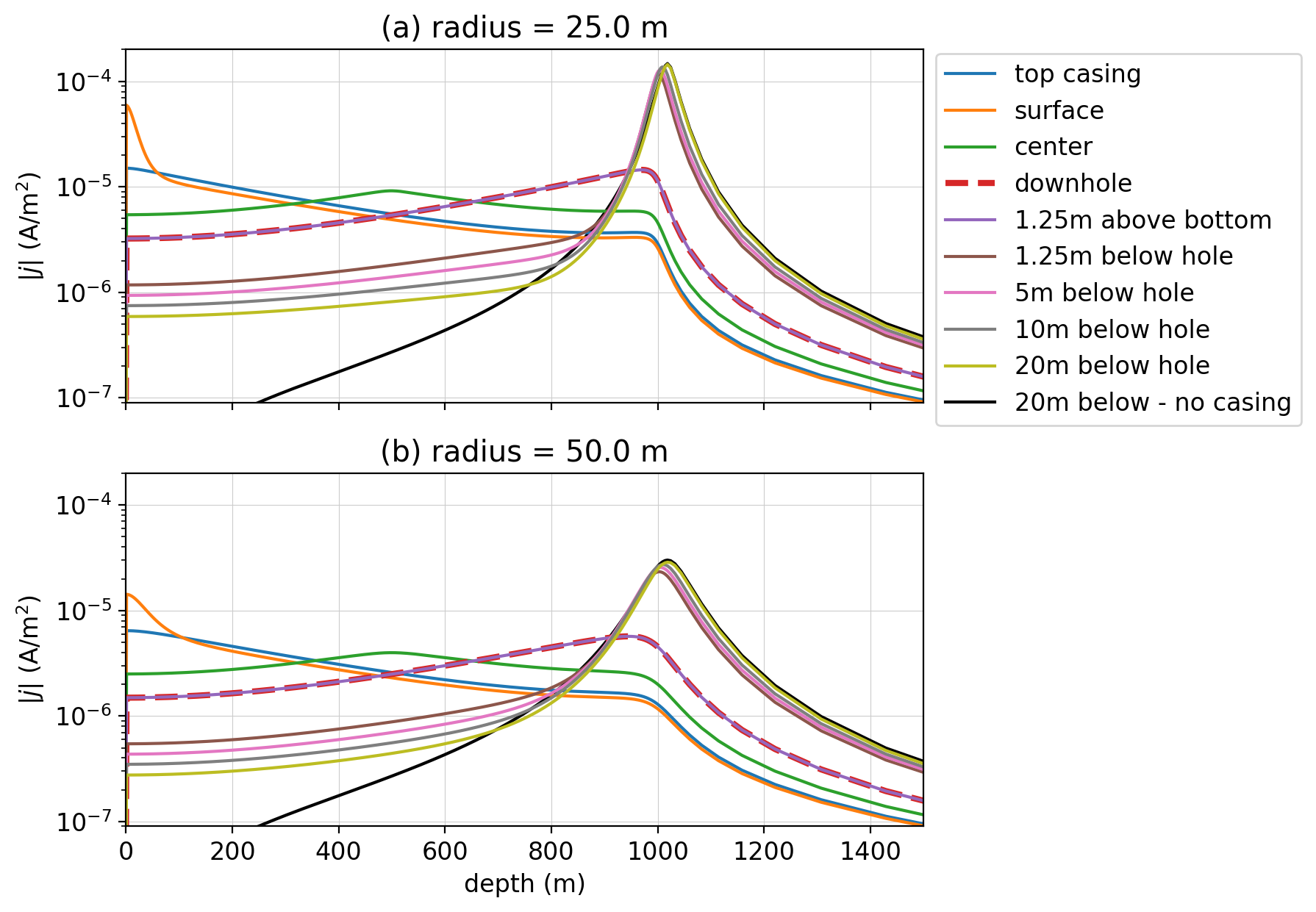}
    \end{center}
\caption{
    Total current density along a vertical line offset (a) 25 m and (b) 50 m
    from the axis of the casing, which extends
    from the surface (0 m) to 1000 m depth.
    The electrode locations correspond to those shown in Figure \ref{fig:electrode_location}.
    For reference, a simulation with an electrode 20m below the casing when there is no casing present
    is shown in black.
}
\label{fig:electrode_location_currents}
\end{figure}

To assess the ability of each electrode configuration to excite a geologic target of interest, we will examine the current density in the formation. In Figure \ref{fig:electrode_location_currents}, we have plotted the amplitude of the current density along a vertical line (a) 25 m and (b) 50 m radially offset from the well. In terms of survey design, we wish to choose a source location that maximizes the total current density within the depth region of interest. If the target is near the surface, we choose an electrode which is connected to the top of the casing, or near the casing at the surface. Interestingly, at depth, there is little distinction between these two scenarios. This has been similarly noted by \cite{Patzer2017} (Figure 10 in particular). Thus, if one is limited to deploying electrodes at the surface, and for practical purposes, connecting infrastructure to the well-head presents a challenge, then grounding the electrode near the well still results in a survey that benefits from the well acting as a high-conductivity pathway to help deliver current to depth. If the aim however, is to excite a deeper target, we see that positioning the electrode downhole can significantly increase the current density delivered to that depth. For example, if we have a target near 500 m depth, positioning the electrode near that depth nearly doubles the current density as compared to an electrode at the surface. If a target is near the end of the well, between 800 m and 1000 m depth, then positioning an electrode near the end of the well triples the current density. This effect will be amplified if the well is lengthened, since we observe exponential decay of the currents carried along according to the conduction length (equation \ref{eq:conduction_length}).

\cite{Kaufman1990} pointed out that the difference in the distribution of currents between a survey where an electrode is positioned along the axis of the casing and one in which the electrode is coupled to the casing is localized near the electrode. Hence, whether the electrode is coupled to the casing or not is not an important distinction at the scales we consider for geophysical imaging. We can test this numerically by comparing the currents arising from the electrode which is connected to the casing 5m above the bottom of the casing (red in Figures \ref{fig:electrode_location} and \ref{fig:electrode_location_currents}), and the electrode positioned along the axis of the casing 1.25 m above the bottom of the casing (purple in Figures \ref{fig:electrode_location} and \ref{fig:electrode_location_currents}). Indeed, we see that the red and purple lines overlap for all offsets in Figure \ref{fig:electrode_location_currents}, indicating that both situations result in the same distribution of currents within the formation.

For electrodes beneath the casing, the distribution of currents is significantly different. For electrodes 1.25 m, 5 m, 10 m and 20 m below the casing, we see that within a $\sim$ 100 m above and below the electrode location, the currents are nearly symmetric, following the expected response of a point source. We have included a simulation with the electrode 20 m below the pipe when there is no casing present; this is shown in black in Figure \ref{fig:electrode_location_currents}. The main difference between the distribution of currents for each of these scenarios is the reduction in current density in the top 1000 m, with increasing electrode depth; as the electrode is moved deeper, less current is channeled into the casing. \cite{Schenkel1990} noted that for electrodes positioned beneath a well, if the electrode is more than 100 casing diameters beneath the casing, then the casing has little impact on the fields below or far from the pipe. The current is much more localized if the electrode is beneath the casing, and thus if a target is beneath or very near the end of the well, then it is advantageous to position the electrode beneath the well.

Not surprisingly, if the source electrode can be positioned near the depth region of interest, the current density delivered to that region is larger. Numerical experiments show that the position of the return electrode makes minimal impact on the currents at depth. However, if the return electrode is within 10’s of meters of the well, the near surface currents are significantly altered. This is consistent with our observations in Section \ref{sec:casing_integrity}, where we showed that the return electrode location has little impact on the magnitude of the secondary signals, but its position alters the geometry of the source fields and this can be used to reduce coupling of receivers to the primary field.

\subsection{Target properties}
The physical property contrast between the target and the background, the target's geometry, and its proximity to the well, all influence our ability to observe its impact on the data we measure. The purpose of this section is to explore the impact of these factors on the excitation and detection of the target. In the first example, we examine the role of the conductivity of a cylindrical target which is in contact with  the well. The second example is again a cylindrically symmetric co-axial disc target but there is a gap between the casing and the target. For the following numerical simulations, the 3D cylindrical code is used.
\subsubsection{Target in contact with the well}
First, we consider a cylindrical target that is in contact with the well. \cite{Schenkel1994} examined such a scenario for a conductive target (e.g. a steam injection or water flood) in a mise-\`a-la-masse type experiment where a source electrode is connected to the casing at the same depth as the center of the target. They considered a cross-well experiment with potential electrodes in an offset, uncased well, and compared two scenarios for the source well: one in which the source well is an open-hole, and the second in which it was cased. They demonstrated that the casing enhances the response, and thus the data sensitivity to the target, as compared to an experiment where current is injected directly into the target and no casing is present. In this example, we build upon those findings and examine the role of the conductivity of the target on our ability to excite it as well as the impact on the data if the target is not directly in contact with the well.

The model we use is a 1 km casing in a half-space with a target. The target extends 25 m vertically and has a 25 m radius and the depth to its top is 900 m. The model is cylindrically symmetric and thus we expect that the secondary electric field at the surface due to the target will be purely radial. As such, we apply the learnings from the casing integrity example and use the return electrode to reduce coupling with the primary field along a line perpendicular to the source. We position the return electrode 500 m from the well-head and we compare both top-casing and down-hole source electrode locations.

\begin{figure}
    \begin{center}
    \includegraphics[width=\textwidth]{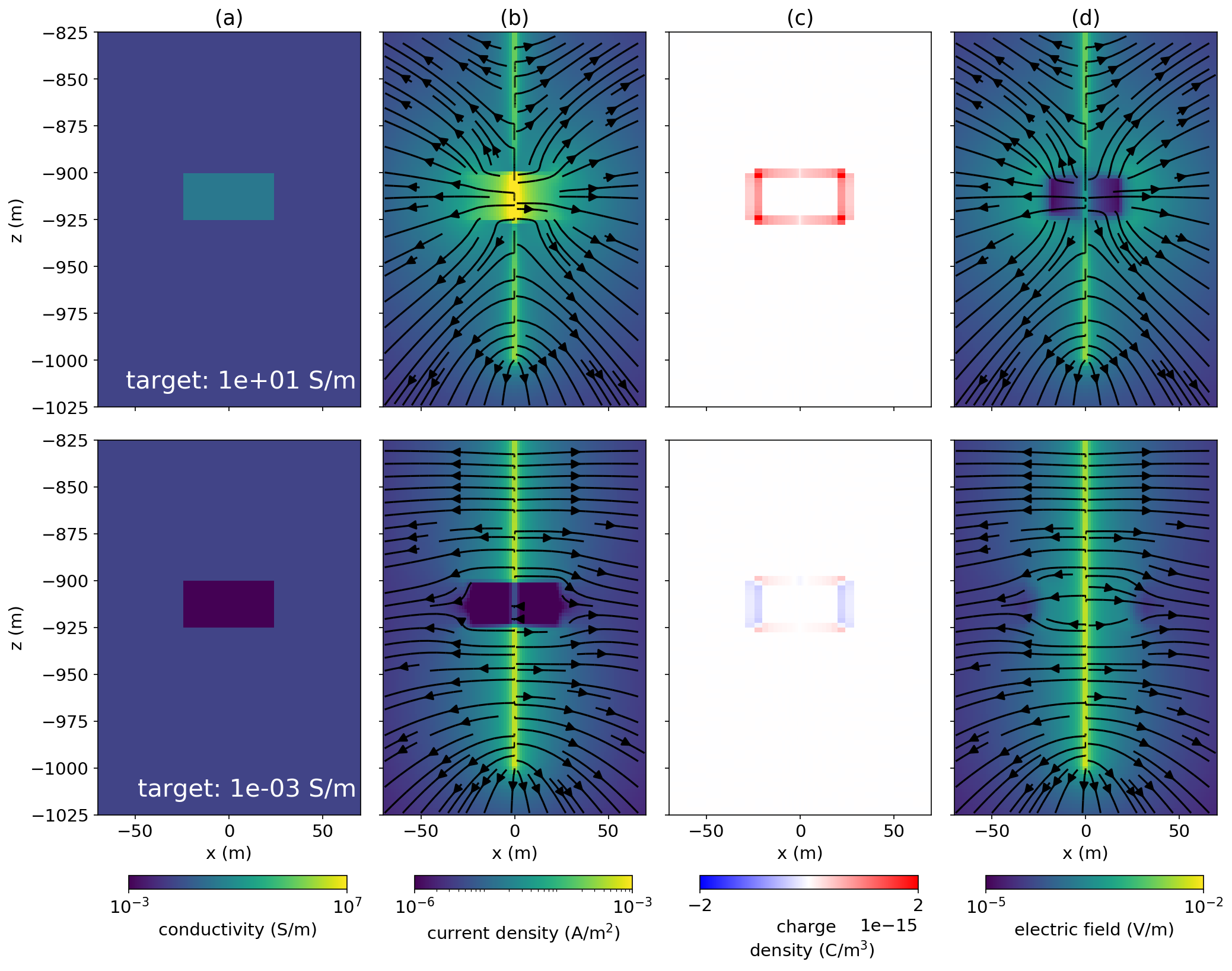}
    \end{center}
\caption{
    Cross section showing: (a) electrical conductivity, (b) current density, (c) charge density, and
    (d) electric field for a DC resistivity experiment with a conductive target (top) and a resistive target
    (bottom). The positive electrode is positioned in the casing at the 912.5m depth.
    The casing is shown by the black line that extends to 1km
    depth in panel (a).
}
\label{fig:target_physics}
\end{figure}

\begin{figure}
    \begin{center}
    \includegraphics[width=\textwidth]{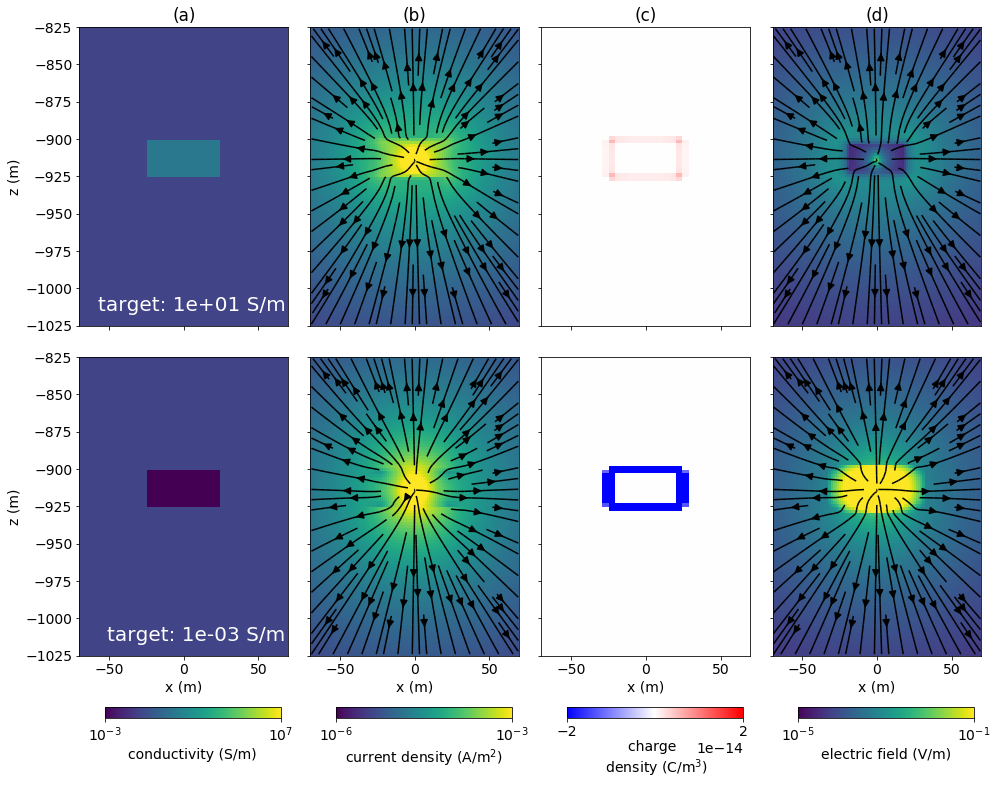}
    \end{center}
\caption{
    Cross section showing: (a) electrical conductivity, (b) current density, (c) charge density, and
    (d) electric field for a DC resistivity experiment with a conductive target (top) and a resistive target
    (bottom). The positive electrode is positioned at 912.5m depth.
    No casing is included in this simulation. Note that the colorbars for the charge density (c) and electric field (d)
    are different than those used in Figure \ref{fig:target_physics}. For the resistive target, the colorbar is saturated,
    the charge density over the resistive target is on the order of $10^{-13}$ C/m$^3$.
}
\label{fig:uncased_target_physics}
\end{figure}

We begin by examining the physical behavior governing the DC response of a conductive and resistive target. Figure \ref{fig:target_physics} shows the (a) conductivity model, and resultant: (b) current density, (c) charge density, and (d) electric fields for a conductive target ($10$ S/m, top row) and a resistive target ($10^{-3}$ S/m, bottom row) in a down-hole experiment where the source electrode is positioned at the center of the target. The extent of the steel-cased well is noted by the vertical black line in panel (a). For the conductive target, we see an accumulation of positive charges along the radial and vertical boundaries of the target. This is consistent with currents that exit a conductor into a more resistive background. The physical response is more complicated when the target is resistive. Intuitively, one might expect that conversely to the conductor, we have a build up of negative charges on the boundary as currents exit a resistor into a more conductive background. This is what would be observed in a traditional mise-\`a-la-masse experiment, where a point source is positioned within the target (Figure \ref{fig:uncased_target_physics}). However, when the casing is present, there is an accumulation of positive charges on the top and bottom boundaries of the target. Currents leak-off along the entire length of the casing, and some of those that leak-off above and below the target are deflected into the target. As a result, there is an accumulation of positive charge on the resistive target. This asymmetry between conductive and resistive targets is not intuitive and demonstrates the power of numerical modelling for understanding the physical responses.

\begin{figure}
    \begin{center}
    \includegraphics[width=\textwidth]{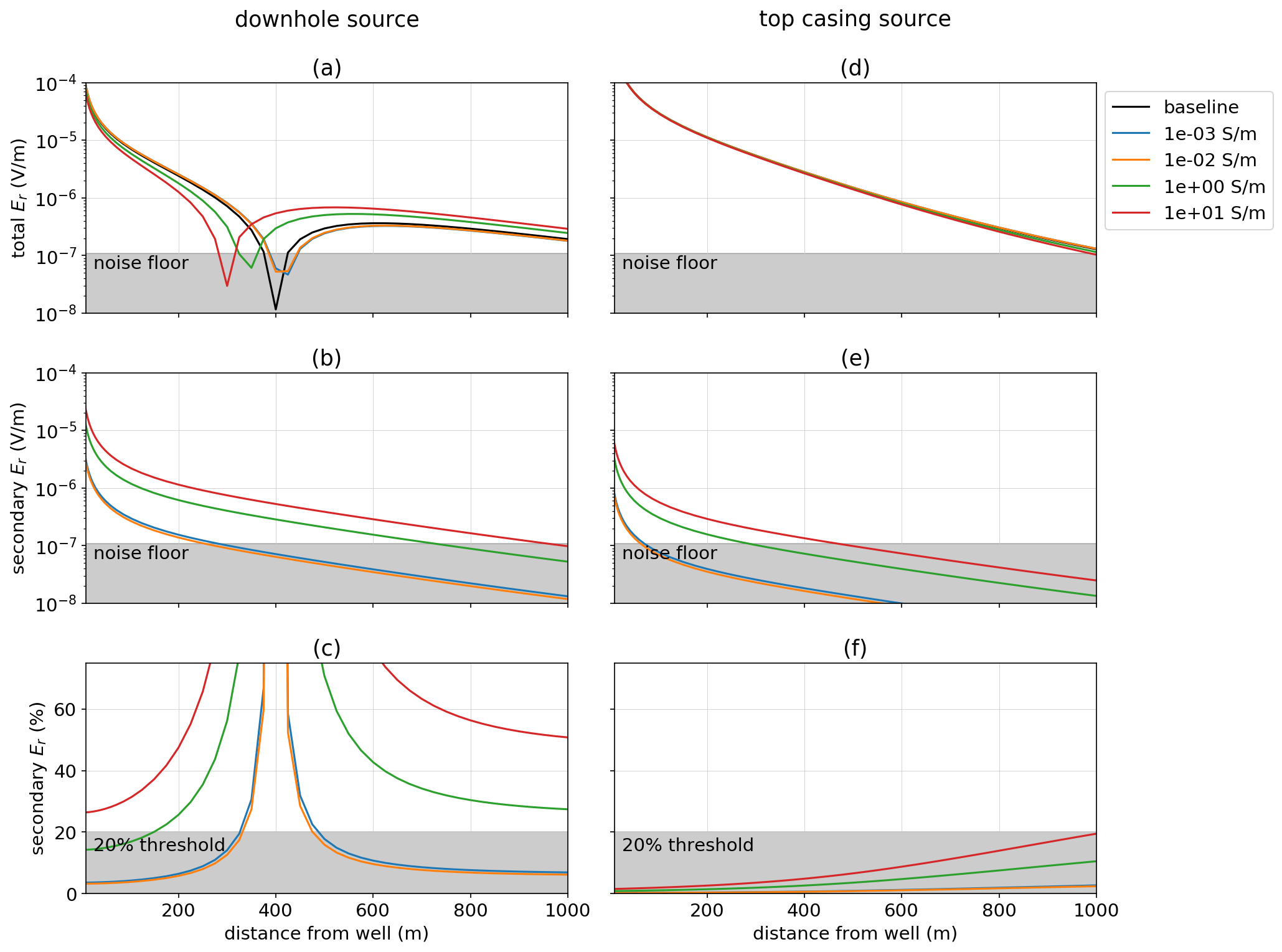}
    \end{center}
\caption{
    Radial electric field at the surface as the conductivity of a cylindrical target, in contact with the well,
    is varied. The target has a radius of 25m and extends in depth from 900m to 925m. The return electrode
    is on the surface, 500m from the well and data are measured along a line perpendicular to the source.
    The panels on the left show
    (a) the total electric field, (b) the secondary electric field with respect to a primary that does not include the target,
    and (c) the secondary electric field as a percentage of the primary for a survey in which the positive electrode is
    positioned downhole at 912.5m depth. The panels on the right similarly show (d) the total electric field, (e) the
    secondary electric field, and (f) the secondary electric field as a percentage of the primary for a top-casing experiment.
}
\label{fig:target_electric_fields}
\end{figure}

In a DC experiment, the electric field response we measure is a result of the distribution of charges within the domain. As a metric for quantifying excitation, we integrate the secondary charge over this depth interval containing the target. In Table \ref{tab:target_charge}, we show the secondary charge integrated over the depth interval containing the target; the secondary charge on the casing within this region is included in the calculation. To examine how the charge relates to the electric field data, we have plotted (a) total radial electric field, (b) secondary radial electric field (with respect to a primary that includes the casing in a halfspace), and (c) the secondary radial electric field as a percentage of the primary for a down-hole source and similarly for a top-casing source (d, e, f) in Figure \ref{fig:target_electric_fields}. We have adopted the same noise floor and percent threshold as in the casing integrity examples ($10^{-7}$ V/m and $20\%$, respectively). For time-lapse surveys where a baseline survey has been taken and the background is well-characterized, this threshold could likely be reduced. The black line in panels (a) and (d) corresponds to the baseline model in which no target is present; each of the colored lines corresponds to a different target conductivity as indicated in the legend.

First, we examine the impact of the conductivity of the target and notice that there is an asymmetry between secondary charge on conductive targets and resistive targets. For a 1 S/m target, which is one order of magnitude more conductive than the background, the integrated secondary charge is $1.75 \times 10^{-11}$ C, while for a $1\times10^{-2}$ S/m target, which is one order of magnitude more resistive than the background, the integrated secondary charge is $-3.82 \times 10^{-12}$ C for the downhole casing experiment. Thus, there is a factor of 4.6 between the magnitude of the secondary charge for these targets; this is equivalent to the ratio we see between the secondary electric field measurements at the surface observed in Figure \ref{fig:target_electric_fields}b. When also considering the influence of the primary electric field on our ability to detect a target, we see that for a down-hole casing experiment, the conductive targets are detectable; they both have a significant region where the secondary is above the noise floor and the secondary comprises a significant percentage of the primary. The resistive targets, however, are not. Although within 200m of the well, the secondary signal is above the noise floor, this also corresponds to where the primary field is large; the percent threshold would need to be reduced to less than 5\% in order to have confidence in the signals due to the resistive targets.

When comparing the downhole source to the top-casing source experiments for a fixed conductivity, there is a factor of 3.9 between the integrated secondary charge shown in \ref{tab:target_charge}; this is reflected in the secondary electric field data in Figure \ref{fig:target_electric_fields} (b) \& (e).  For the top-casing experiment, none of the targets is detectable. There are two factors that make this a more challenging experiment than the downhole scenario: (1) less current is available to excite the target, as reflected in Table \ref{tab:target_charge} and (2) the primary field is stronger at the receivers (200m from the well the primary field has an amplitude of $10^{-5}$ V/m, while for the down-hole source experiment, the primary has an amplitude of $2 \times 10^{-6}$ V/m). Addressing the excitation of the target requires that the source electrode be positioned downhole, closer to the target. The second point may be overcome if receivers can be positioned closer to the target, for example within an adjacent borehole.

In summary, the integrated secondary charge provides a metric for a survey's ability to excite a target, and shows that conductive targets are easier to excite than resistive targets. As expected, if the source electrode can be positioned near the target, excitation is enhanced. This also has the added benefit of reducing the strength of the primary electric field at the surface, as compared to a top-casing survey; this  increases the potential for detecting a target with surface-based receivers. In the next section, we examine the significance of the electrical connection between the casing and the target.

\subsubsection{Target not in contact with the well}

\begin{figure}
    \begin{center}
    \includegraphics[width=\textwidth]{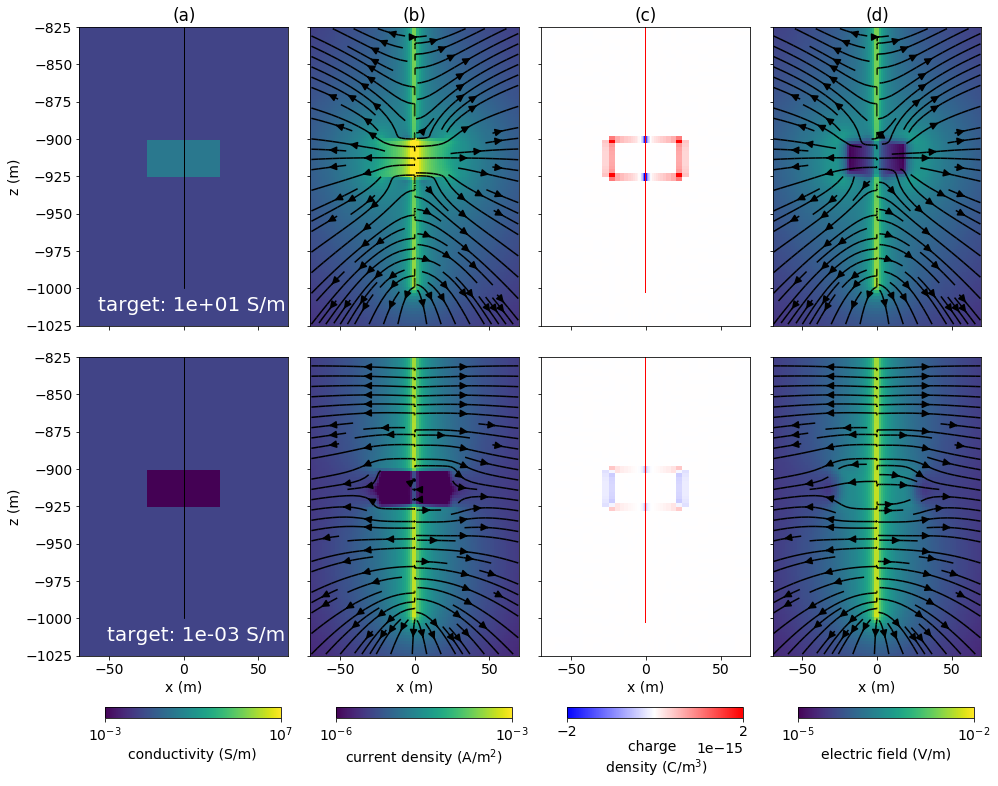}
    \end{center}
\caption{
    Cross section showing: (a) electrical conductivity, (b) current density, (c) charge density, and
    (d) electric field for a DC resistivity experiment with a conductive target (top)
    and a resistive target (bottom) which is not in contact with the well.
    The positive electrode is positioned in the casing at the 912.5m depth.
    The casing is shown by the black line that extends to 1km
    depth in panel (a).
}
\label{fig:offset_target_physics}
\end{figure}

How significant is the electrical connection between the casing and the target for our ability to excite a response? To examine this, we introduce a small gap equal to the thickness of the casing (1cm) between the casing and the target. This has negligible effect on the volume of the target, but it changes the electrical characteristics of the problem. Consider a conductive target; if it is in-contact with the well, we are effectively conducting a mise-\`a-la-masse experiment, and the conductor will have a net positive charge. When the target is isolated from the casing, the total charge on the target must be zero, and thus dipolar effects, in which negative charges build up on the inner interface of the cylinder target and positive charges build up on the outer interface of the target, will be the source of our signal. This is demonstrated in Figure \ref{fig:offset_target_physics}.

\begin{figure}
    \begin{center}
    \includegraphics[width=\textwidth]{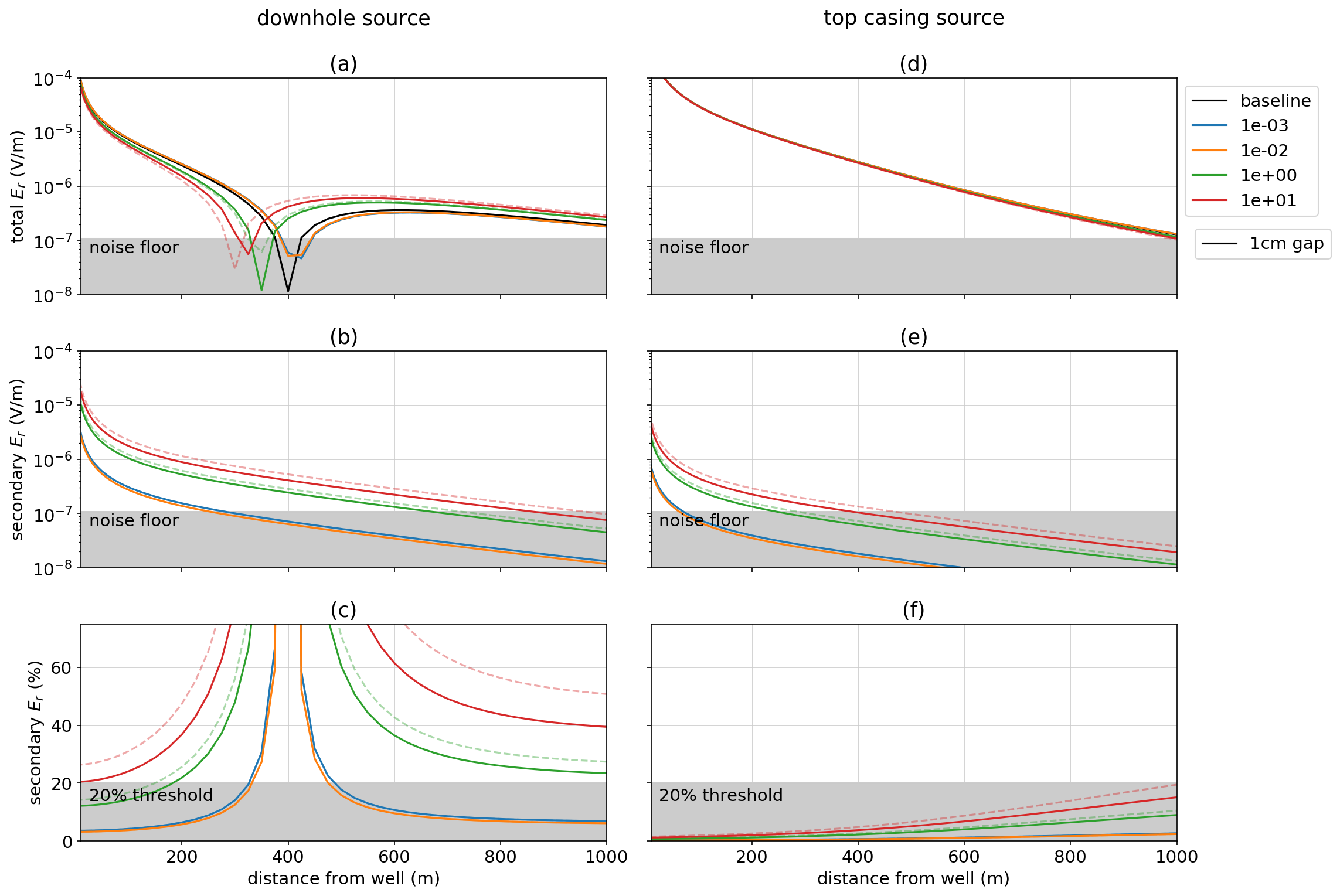}
    \end{center}
\caption{
    Radial electric field at the surface as the conductivity of a cylindrical target, which is not in contact with the well,
    is varied. The target has a radius of 25m and extends in depth from 900m to 925m. The return electrode
    is on the surface, 500m from the well and data are measured along a line perpendicular to the source.
    The panels on the left show
    (a) the total electric field, (b) the secondary electric field with respect to a primary that does not include the target,
    and (c) the secondary electric field as a percentage of the primary for a survey in which the positive electrode is
    positioned downhole at 912.5m depth. The panels on the right similarly show (d) the total electric field, (e) the
    secondary electric field, and (f) the secondary electric field as a percentage of the primary for a top-casing experiment.
    The data shown in Figure \ref{fig:target_electric_fields}, for the target in contact with the well,
    are plotted in the dashed, semi-transparent lines for reference.
}
\label{fig:offset_electric_fields}
\end{figure}

The corresponding secondary charge integrated over the target depth and radial electric field data are shown in Table \ref{tab:offset_charge} and Figure \ref{fig:offset_electric_fields}. For comparison, the data resulting from the target in contact with the well are plotted in the dashed, semi-transparent lines. While there is little difference in the integrated secondary charge or the electric field measurements for the resistive targets, we see that there is a factor of 1.3 difference (i.e. 30\%) between the integrated secondary charges and correspondingly, the secondary electric fields, from a 10 S/m target in contact with the well versus not. Similarly, there is a factor of 1.2 (20\%) between a 1 S/m target in contact with the well versus not for both the downhole and top-casing sources. Increasing the gap between the target and the casing decreases the integrated charge and correspondingly reduces the secondary electric field at the surface. The integrated secondary charge for a 10 S/m target with a 10cm gap between the target and casing in a downhole source experiment is $1.7 \times 10^{-11}$ C, which is a factor of 2.2 smaller than the connected target; correspondingly the electric field data at the surface are reduced by a factor of 2.2 as compared to the connected target. Thus, a direct, electrical connection between the target and the well in which we connect the source is preferable for exciting and detecting conductive targets.

Designing a survey for a specific setting may require incorporation of 3D geologic structures and may include inversions to examine a survey's ability to recover a target. In this case, it is desirable to have a coarse-scale representation the steel-cased well on the simulation mesh. This is the topic of the next section.

\section{Coarse-scale approximations of the well}
\label{sec:approximating_wells}
When approaching the inverse problem, many forward simulations are required, and typically, a 3D cartesian mesh, with cells that vary on the length scales of the geology, is desired. Thus, rather than performing a fine-scale simulation of the steel-cased well, we may wish to represent the well on a coarse mesh. In the literature, two common approaches arise: the first approximates the well as some form of ``equivalent source,'' such as a charge distribution (e.g. \cite{Weiss2016}); the second approach represents the well as a conductivity feature on the coarse-mesh (e.g. \cite{Swidinsky2013, Um2015, Yang2016, Kohnke2017, Puzyrev2017}, among others). Here, we will focus our attention to the second approach, noting that a charge distribution along the length of the well can be computed with the 2D or 3D cylindrical code described in \cite{Heagy2018a}. Within the literature, there is disagreement among approaches for selecting the conductivity of the coarse-scale feature approximating the well. For example, \cite{Um2015} replaces the fluid-filled cylinder with a solid rod having the same conductivity as the casing, arguing that it is the contrast between the conductivity of the well and the conductivity of the surrounding geology that is the most important factor; \cite{Puzyrev2017} also adopts this approach. Other authors have opted to preserve the cross-sectional conductance of the well \citep{Swidinsky2013, Kohnke2017}; this is consistent with the transmission-line model of the well discussed in \cite{Kaufman1990}. The aim of this section is to analyze these approaches.
\subsection{Replacing a hollow-cased well with a solid cylinder}
We consider a steel-cased well with a conductivity of $5\times10^6$ S/m that is embedded in a 0.1 S/m halfspace; the conductivity of the material that fills the well is the same as the background. The well has an outer diameter of 10cm and a thickness of 1cm, and we will vary its length. We will perform a top-casing experiment, where the positive electrode is connected to the casing at the surface. The return electrode is positioned 8km away, and a cylindrically symmetric mesh is used in the simulations. We examine approximations that treat the casing as a solid cylinder with the same outer-diameter as the true, hollow-cased well.

\begin{figure}
    \begin{center}
    \includegraphics[width=\textwidth]{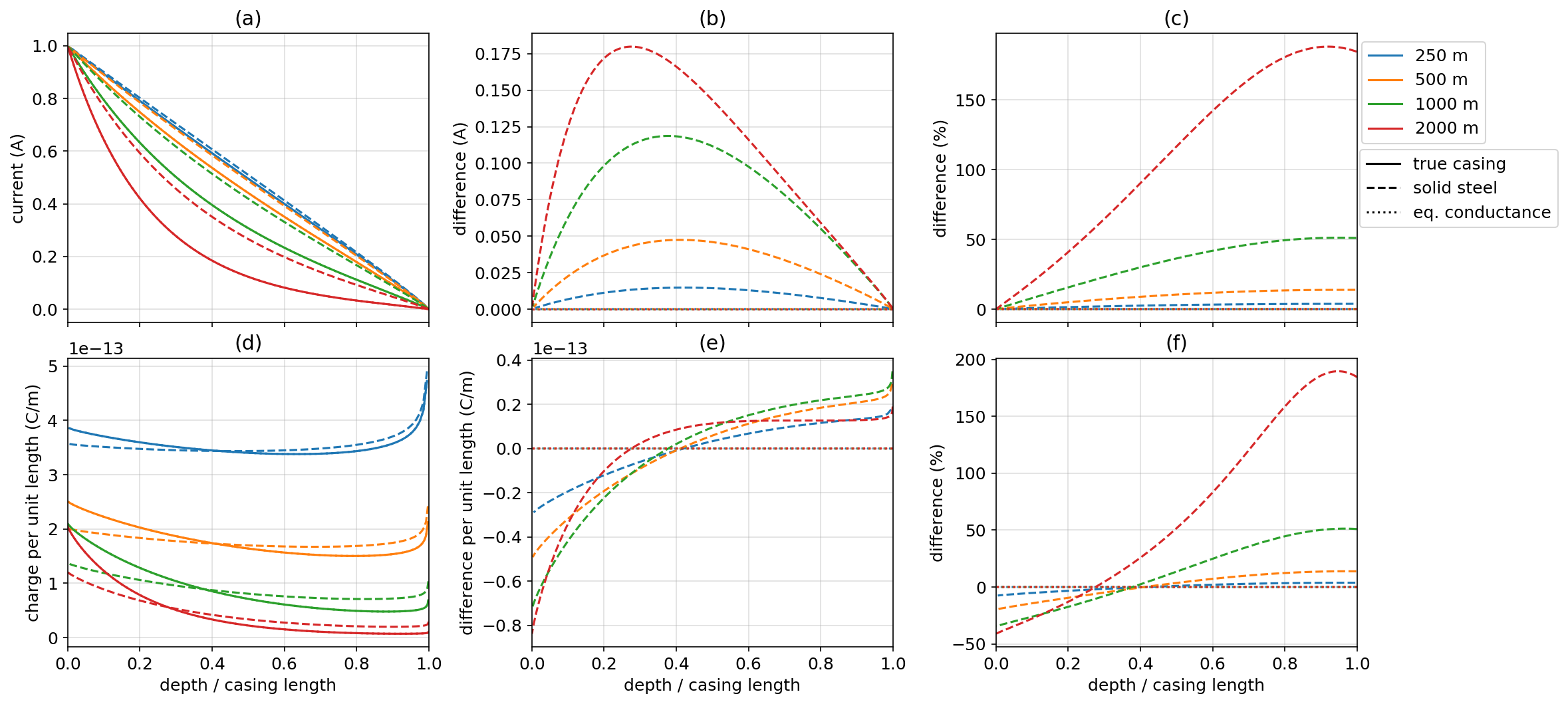}
    \end{center}
\caption{
    Currents (top row) and charges (bottom row) along the length of
    a hollow steel-cased well (solid lines), solid cylinder with
    conductivity equal to that of the steel-cased well (dashed-lines),
    and a solid cylinder with a conductivity such that the product of the
    conductivity and the cross sectional area of the cylinder is equal to that
    of the hollow-pipe (dotted lines). Each of the line-colors corresponds to a
    different casing length, as indicated in the legend.
    In (a), we show the vertical current in the casing,
    (b) shows the difference from the true, hollow-cased well
    in the vertical current within the casing, and (c) shows that difference as a percentage
    of the true currents. In (d), we show the charge per unit length along the casing, (e)
    shows the difference from the true, hollow-cased well and (e) shows that differences as
    a percentage of the true charge distribution.
    The x-axis on all plots is depth normalized by the length of the casing.
}
\label{fig:approximating_wells_currents_charges}
\end{figure}

The distribution of charges, or equivalently, the current in the casing, is the source of the electric response of the casing. Thus to judge if two models of the casing are ``equivalent'', we examine the current and charges as a function of depth. In Figure \ref{fig:approximating_wells_currents_charges}, we have plotted the vertical current and charges along the casing for the true, hollow cased well (solid), solid cylinder with conductivity equal to that of the casing, $5 \times 10^6$ S/m (dashed), and solid cylinder with a conductivity that preserves the product of the conductivity and the cross-sectional area of the conductor, $1.8 \times 10^6$ (dotted), for four different casing lengths, each indicated by a different color. Figure \ref{fig:approximating_wells_currents_charges} shows: (a) the vertical current along the casing, (b) the difference in current between the approximate model and the true model, (c) that difference as a percentage of the true solution (d) the charge per unit length, (e) difference in charge per unit length and (f) difference in charge per unit length as a percentage of the true solution.

For short wells, we see that the current decays linearly and that the charge distribution is nearly uniform above the end of the well, while for longer wells, the decay of the current is exponential in nature, as is the charge distribution. This behavior is consistent with that predicted by the transmission line solution described in \cite{Kaufman1993}. \cite{Kaufman1993} showed that the transition between the linear decay of currents and the exponential decay of currents is controlled by three factors: the cross sectional conductance of the well, the resistivity of the surrounding formation, and the length of the well. \cite{Schenkel1991} similarly summarized this behavior in the definition of the conduction length (equation \ref{eq:conduction_length}), which is the length over which the currents in the casing have decayed by a factor of $1/e$. For sufficiently conductive and short wells (e.g. $L_c / \delta \ll 1$, where $L_c$ is the length of the casing), the current decay is linear and independent of the conductivity, whereas for longer wells, ($L_c / \delta \gg 1$), the rate of decay of the currents is controlled by the conduction length (see equations 45 and 53 in \cite{Kaufman1993}).

\begin{figure}
    \begin{center}
    \includegraphics[width=\textwidth]{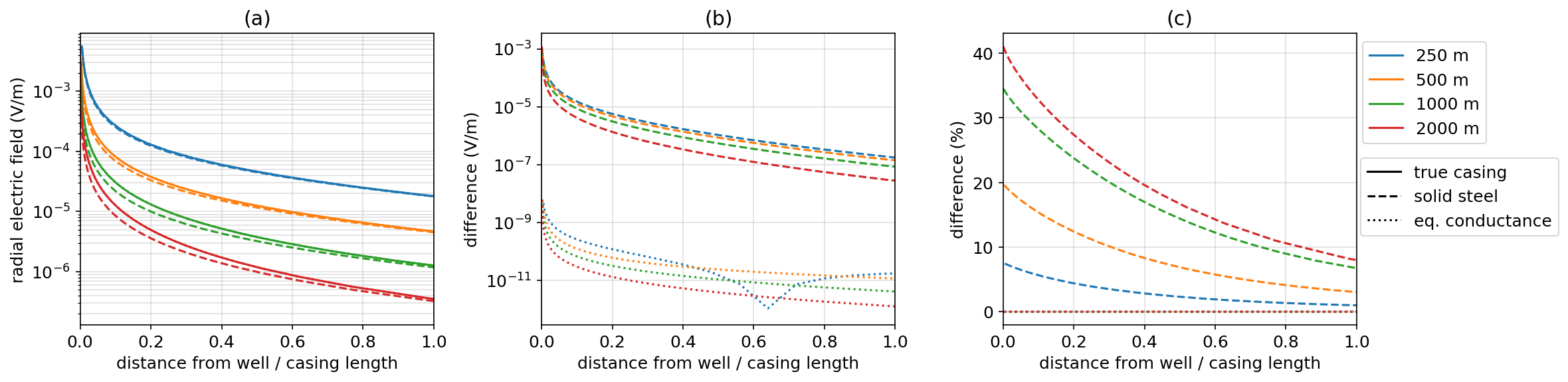}
    \end{center}
\caption{
    Radial electric field measured at the surface for a model of
    a hollow steel-cased well (solid lines), a solid cylinder with
    conductivity equal to that of the steel-cased well (dashed-lines),
    and a solid cylinder with a conductivity such that the product of the
    conductivity and the cross sectional area of the cylinder is equal to that
    of the hollow-pipe (dotted lines). Each of the line-colors corresponds to a
    different casing length, as indicated in the legend.
    In (a), we show the total radial electric field,
    (b) shows the difference in electric field from that due to the true, hollow-cased well,
    and (c) shows that difference as a percentage
    of the true electric fields.
    The x-axis on all plots is distance from the well normalized by the length of the casing.
}
\label{fig:approximating_wells_electric_fields}
\end{figure}

In preserving the cross-sectional conductance, we see that the difference in currents and charges along the length of the well is negligible;  the maximum difference in currents for the 2000m long well which has equivalent cross-sectional conductance is $7\times10^{-7}$ A as compared to the difference of 0.18 A when using the conductivity of the casing. This difference is important as it changes how much current is available to excite a target at depth. For a 2000m long well, the current is overestimated by $> 150\%$ if the well is replaced by a solid cylinder with the same conductivity of the steel-cased well. It also changes the distribution of charges and thus the electric field due to the well. Figure \ref{fig:approximating_wells_currents_charges}e shows us that the extra conductance introduced when approximating the well using the conductivity equal to the casing results in a secondary dipolar charge on the casing. This in turn reduces the electric field we observe at the surface, as shown in Figure \ref{fig:approximating_wells_electric_fields}. For a long well, the difference can be as large as 40\% near the well.

The numerical time-domain EM experiment used in \cite{Um2015} to demonstrate the approximation of the well by a solid, conductive rod having the same conductivity as the steel-cased well used a 200m long well with a thickness of 12.223mm, outer diameter of 135mm, conductivity of $10^{6}$ S/m in 0.033 S/m half-space. The conduction length of this well is 560m; this is more than twice the length of the well. Therefore, the behavior of the currents falls into the linear regime, where the decay of currents is mostly independent of the conductivity, and thus the difference between using the conductivity of the casing or preserving cross-sectional conductance is less significant. However, if longer wells such as those typically employed in hydrocarbon settings, are considered, the behavior of the currents and charges depends upon the conductance of the casing, and thus that is the quantity that should be conserved in an approximation of the hollow-cased well by a solid rod.

\begin{figure}
    \begin{center}
    \includegraphics[width=\textwidth]{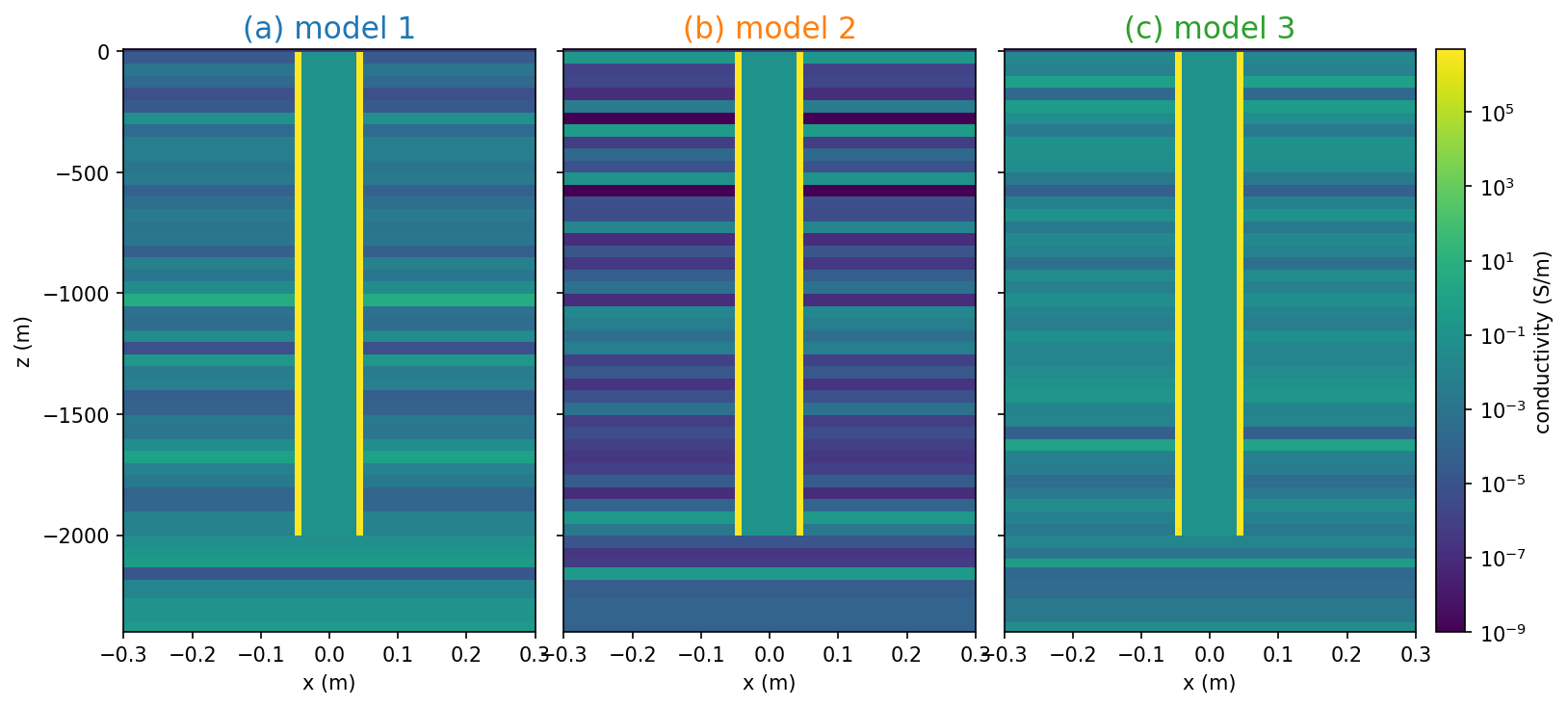}
    \end{center}
\caption{
    Three realizations of a 2km long casing in a layered background, where the conductivity of the
    layers is assigned randomly. Each layer is 50m thick, and the mean conductivity of the background
    is 0.1 S/m. The color of the title corresponds to the plots of the currents and charges in Figure
    \ref{fig:approximating_wells_currents_charges_random}
}
\label{fig:random_layers}
\end{figure}

In order to confirm that this conclusion is valid for variable geology, we have included a simulation with a 2km long casing in a layered background. Each layer is 50m thick and the conductivity was assigned randomly; three instances are included, as shown in Figure \ref{fig:random_layers}. The mean of the background conductivity is 0.1 S/m for each of the models.

\begin{figure}
    \begin{center}
    \includegraphics[width=\textwidth]{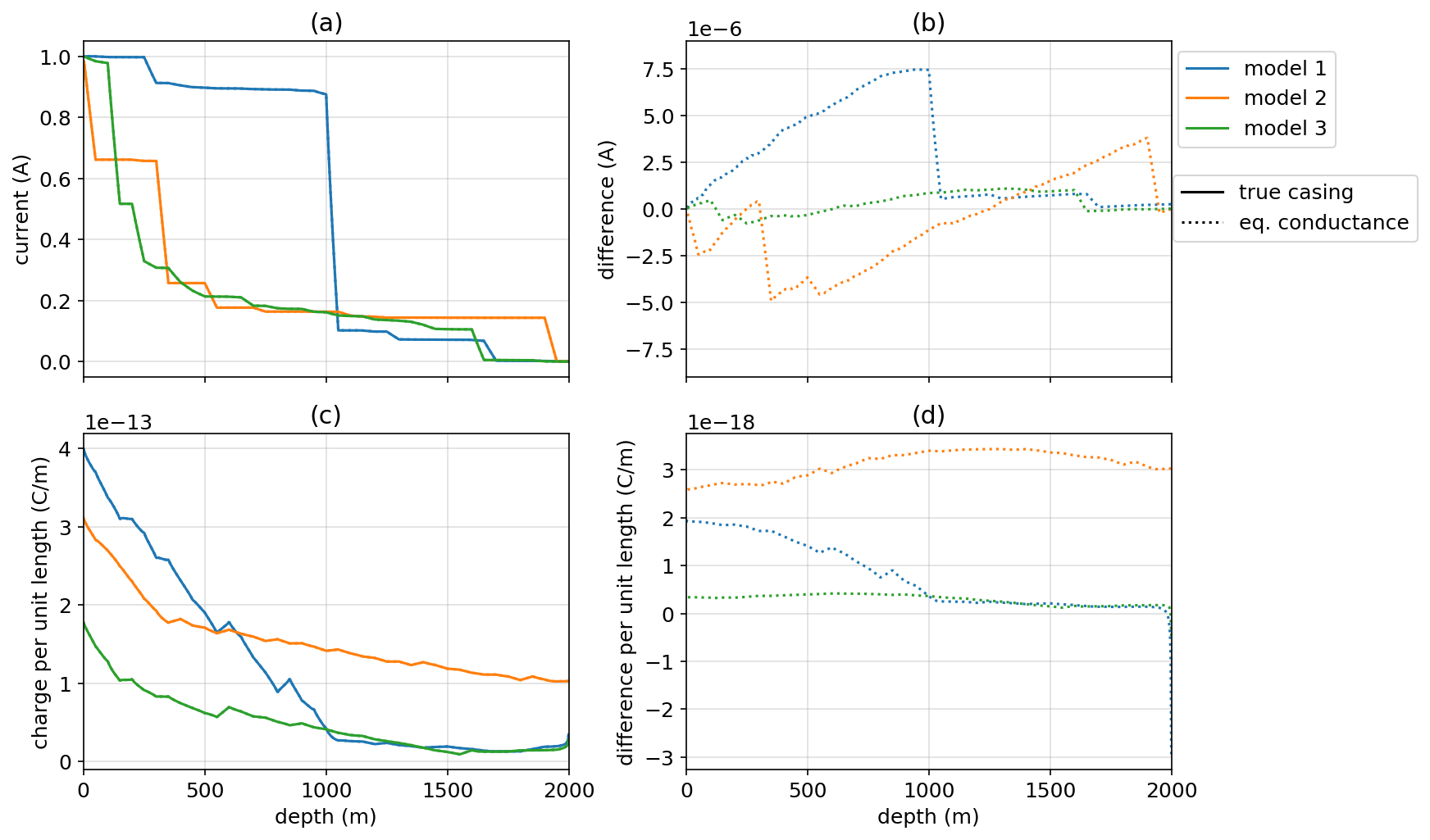}
    \end{center}
\caption{
    (a) Total vertical current through the casing for the three layered-earth models shown in
    Figure \ref{fig:random_layers}. The solid lines indicate the response of the true, hollow steel cased-well
    and the dotted lines indicate the response of a solid cylinder having the same cross-sectional conductance
    as the hollow well. (b) Difference between the currents along the casing in the solid well approximation
    and the true, hollow well.
    (c) Charge per unit length for each of the models. (d) Difference in charge per unit length between the
    true model of the casing and the approximation which preserves cross-sectional conductance.
}
\label{fig:approximating_wells_currents_charges_random}
\end{figure}

The currents and charges along the length of the well for the true model, and a model approximating the well as a solid cylinder with equal cross-sectional conductance, are shown in Figure \ref{fig:approximating_wells_currents_charges_random}. For all of the models shown, the difference in both the casing currents and the charges are 5 orders of magnitude less than the amplitude of the total currents and charges; thus we conclude that approximating a hollow cylindrical steel casing by a solid cylinder with a conductivity that preserves cross-sectional conductance is valid for models with variable geology.

\subsection{Cartesian grid}
In the previous section, we showed that a hollow, cylindrical steel-cased well can be approximated by a solid cylinder with equal cross-sectional conductance. In this section, we move to a coarser, cartesian mesh, such as might be employed in settings with multiple boreholes. We examine a simple approximation of a steel cased well on a cartesian grid. We employ 4 tensor meshes, each with progressively larger cell widths for the finest cells that capture the casing. On each of the cartesian meshes, we approximate the casing by preserving the product of the conductivity and the cross sectional area on the mesh. For comparison, we run a fine-scale simulation on a 3D cylindrical mesh that accurately discretizes the casing; it uses 4 cells across the casing-wall. The casing model is similar to that used in previous examples: it is 1km long, has an outer diameter of 10cm, a thickness of 1 cm, and is embedded in a 0.1 S/m half-space. The positive electrode is connected to the top of the casing and a return electrode is positioned 1km from the well-head. Table \ref{tab:cartesian_simulation} summarizes the number of cells in each mesh and the computation time for each simulation.

\begin{figure}
    \begin{center}
    \includegraphics[width=\textwidth]{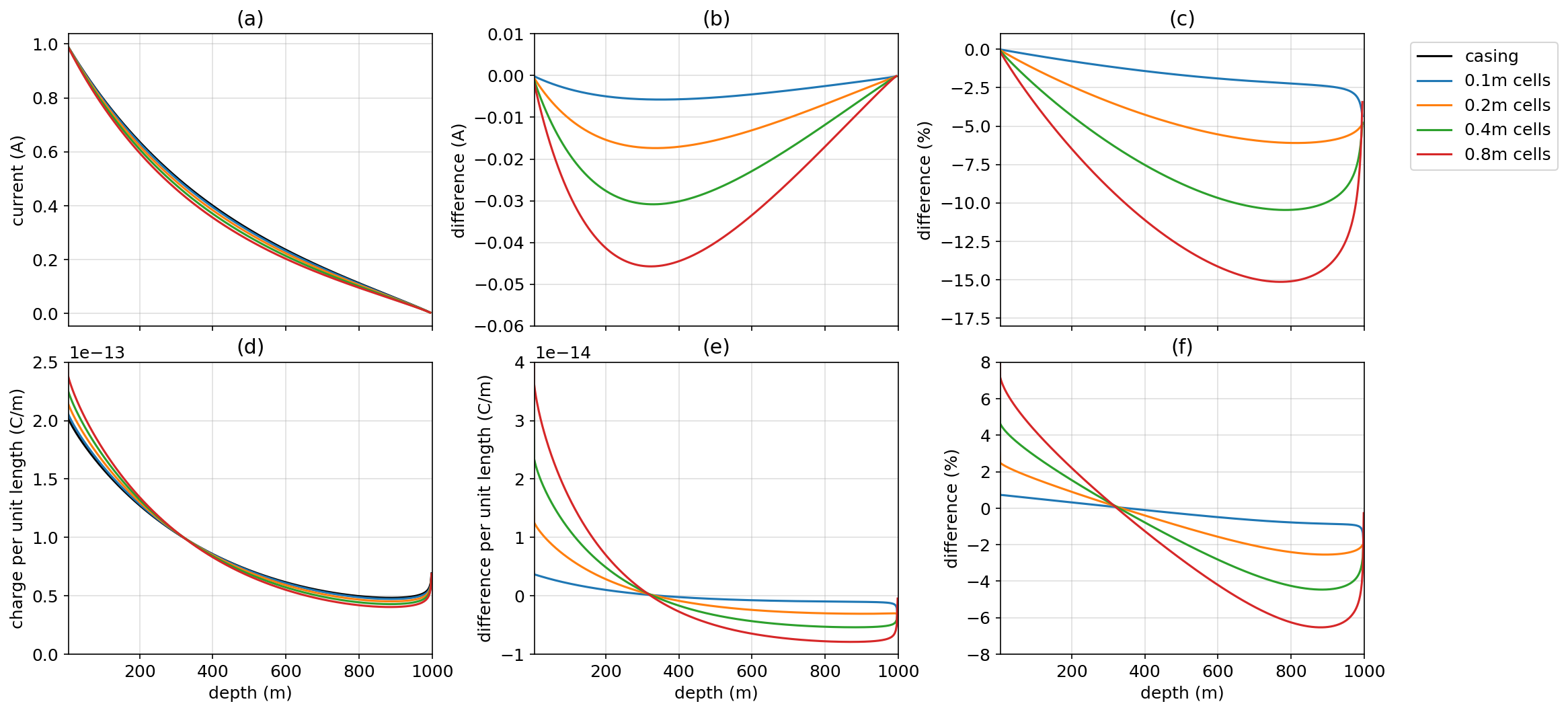}
    \end{center}
\caption{
    Currents (top row) and charges (bottom row) along the length of
    a steel cased well. The ``true'' hollow-cased well is simulated on a
    3D cylindrical mesh and has 4 cells across the width of the casing thickness (black line).
    The colored lines correspond to the currents and charges computed along the
    well represented on a cartesian mesh with cell widths shown in the legend.
    The finest vertical discretization is 2.5m in all simulations. To represent the
    hollow cased well on the cartesian mesh, the cells intersected by the casing are assigned
    a conductivity that preserves the product of the conductivity and cross-sectional area of the well.
    In (a), we show the vertical current in the casing,
    (b) shows the difference from the true, hollow-cased well
    in the vertical current within the casing, and (c) shows that difference as a percentage
    of the true currents. In (d), we show the charge per unit length along the casing, (e)
    shows the difference from the true, hollow-cased well and (e) shows that differences as
    a percentage of the true charge distribution.
}
\label{fig:approximating_wells_cartesian}
\end{figure}

The resultant currents and charge per unit length are shown in Figure \ref{fig:approximating_wells_cartesian}. In the top row, panel (a) shows the total current in a region approximating the well, along with the total current in the ``true'' cylindrical well (black line), (b) shows the difference between the current through the cartesian cells and the true model, and (c) shows the difference as a percentage. Similarly, in the bottom row, we show (d) the charge per unit length along the cylindrical well (black line) and cartesian-prism approximations, (e) the difference in charge per unit length from the charge per unit length on the true cylindrical model, and (f) that difference as a percentage of the charge per unit length on the cylindrical well.

The approximation of the cylindrical well by a rectangular prism with width equal to the diameter of the casing introduces minimal error in the currents and charges computed using a finite volume approach, even though the casing is only captured by one cell across its width. Comparing the current along the length of the well for the 3D cylindrical well and the cartesian simulation with 0.1m cells, we see that the error introduced is $< 2.5\%$ (until the end of the well where the current approaches zero). Similarly, the difference in the charge per unit length is $< \pm 1.25\%$. As successively coarser discretizations are used, accuracy is gradually lost; by doubling the cell sizes to 0.2m, the error in the currents is $6\%$ at its maximum and $< \pm 3\%$ in the charge along the casing. A factor of 8 increase in cell size (0.8m cells) results in a maximum error of $15\%$ in the currents. It is important to note that the forward simulation is conducted using a finite volume approach; other approaches such as finite difference or integral equation approaches may have worse agreement if care is not taken to handle large physical property contrasts, captured by a single cell, in the simulation.
Note that the behavior of the errors depends upon the properties of the casing (e.g. conductivity and length) as well as the conductivity of the background. This might be expected from the description of the casing conduction length (equation \ref{eq:conduction_length}). If the conduction length is large relative to the length of the well, the currents decay linearly, and the geometry and conductivity of the well are less significant in the behavior of the currents. Alternatively, if the conduction length is comparable to the length of the well, the currents decay exponentially with a decay rate that depends on the geometry and conductivity of the well. For example, if the background is more resistive, increasing the contrast between the casing and background, the reduces the errors. Using a background conductivity of $100 \Omega m$, the maximum error introduced in the current is $< 1\%$ with 0.1m cells and $< 2\%$ with 0.8m cells.

Depending on the level of accuracy required in a 3D simulation, there are several strategies that one might take to reduce this error. In some cases, local refinement can be achieved with a tetrahedral mesh, as is often employed when using finite element techniques (e.g. \cite{Weiss2016}), or an OcTree mesh \citep{Haber2007}. Other, more advanced approaches including upscaling and multiscale could also be considered \citep{Haber2014, Caudillo-Mata2017, Caudillo-Mata2017a}. In an upscaling approach, one inverts for a conductivity model, which might be anisotropic, that replicates the physical behavior of interest \citep{Caudillo-Mata2017}. Multiscale techniques translate conductivity information from a fine-scale mesh to a coarse-scale mesh, on which the full simulation is to be solved, using a coarse-to-fine interpolation that is found by solving Maxwell’s equations on the fine mesh locally for each coarse grid cell \citep{Haber2014, Caudillo-Mata2017a}. For treating multiple wells, \cite{Weiss2017} introduces a finite element scheme which allows the user to define conductivities not only as volume-filling cell-centered values, but also as conductive features on the faces, and edges of the mesh. Other approaches in electromagnetics include approximating the well with a series of electric dipoles \citep{Kohnke2017, Patzer2017}.
The 3D cylindrical forward simulation code described in \cite{Heagy2018a} and used in this example can serve a tool for validating and refining an approach to achieve the desired level of accuracy.

\section{Discussion}
The work in this paper is motivated by the increasing use of steel cased wells in geoscience problems, including monitoring applications such as carbon capture and storage and hydraulic fracturing. For geophysical imaging of targets at depth, the wells are beneficial as they can be used to channel currents to depth and enhance signals at the surface for targets that otherwise would be undetectable from a surface-based survey. Additionally, there is interest in considering the casing itself as the target of the geophysical target in casing integrity experiments; here the aim is to detect flaws or breaks in the casing. These applications, coupled with advances in modeling capabilities, open up the potential for advancing the utility of electrical and electromagnetic imaging in settings with metallic-cased wells.

Despite this potential, the reality is that electric fields, especially if measured at the earth’s surface, are small. Secondary fields might only be a few percent of the primary field, and thus too insignificant to reliably detect the target of interest. The success of using a DC or EM survey then depends upon many details that pertain to understanding the basic physics, the effects of parameters of the casing, the background conductivity, location of the current electrodes, and discerning which fields should be measured. DC resistivity is the starting point, as it allows us to examine the currents, charges, and electric fields in the electrostatic limit, prior to introducing inductive effects and the influence of magnetic permeability in an EM signal. Regarding the physics, a DC survey involves attaching a current generator to a conductive medium. This establishes a steady state current; the signal to which we are sensitive is the electric field that arises from charges that accumulate at interfaces separating regions of different conductivity. For this reason, most of our results are first presented as currents and charges.

The large contrasts in physical properties and significant variation in length scales due to long, thin, cylindrical, steel-cased wells prompt a number of questions about how the DC fields behave. In many cases, the finer details about the physical responses has challenged our intuition. With respect to the casing integrity application there were basic questions: how does a flaw in the pipe affect the currents and electric fields measured at the surface? Does the extent of the flaw change our ability to detect it (e.g. if it has a vertical extent of several meters versus a vertical extent of centimeters)? What happens if the flaw only comprises a part of the well, leaving some connectedness in the casing? When considering a geophysical experiment for imaging a target: is there a significant difference in the currents at depth between scenarios where a source electrode is connected to the well-head at the surface and one where the source electrode is offset from the well by a few meters? Can we detect both conductive and resistive targets? What is the physical mechanism which generates the signal in both scenarios? A major goal of the DC survey will be to excite and detect target bodies. For problems, such as CO$_2$ sequestration, enhanced oil recovery, or hydraulic fracturing, the target may or may not be in contact with the well; how significant is an electrical contact between a target and the well in the data we measure at the surface? Looking towards solving inverse problems in settings with steel-cased wells, it is advantageous to reduce the computational cost of the forward simulation because an inversion requires many forward simulations.  Can a coarse-scale approximation of the well be used? What is the correct conductivity needed for substitution?

Some of the above questions have been addressed in theoretical papers extending back a few decades but numerical verification was often limited or carried out with simplifying assumptions. Other questions require the ability to carry out numerical modeling in 2D or 3D environments -- these tools are just now becoming available. Our goal with this paper has been to examine the scientific questions above and to promote insight about the solution by plotting the currents, charges, and electric fields. This analysis has benefited from the ease with which fields, fluxes and charges are readily calculated and visualized within the SimPEG framework, particularly when used in conjunction with Jupyter notebooks. Source codes for all of the examples in this paper are available in the form of Jupyter notebooks at https://github.com/simpeg-research/heagy-2018-dc-casing \citep{Heagy2018b}; our aim in providing these notebooks is to allow readers to reproduce the results shown and also adapt the parameters and extend the analysis to address their questions.

\section{Conclusions}
In this paper we have provided an overview of the fundamental physics governing the behavior of currents, charges, and electric fields DC resistivity experiments with steel-cased wells for both casing integrity experiments and geophysical imaging applications. With respect to casing integrity, we considered a top-casing DC resistivity experiment to detect an impairment in the well. We showed that if a flaw comprises the entire circumference of some depth interval along the casing, then the charges are concentrated in the portion of the well above the flaw, and to first approximation, the charge distribution is equal to that of a well which has been truncated at the depth of the flaw. This excess charge is the source of our signal. As it is cylindrically symmetric, the resultant secondary electric fields due to the flaw are purely radial. In terms of survey-design, we can take advantage of this knowledge and use the return electrode location to reduce coupling with the primary electric field in our data. Our ability to detect a flaw across the entire circumference of the casing depends upon the conductivity of the background and casing, as well as the depth of the flaw. Larger contrasts between the casing and the background (e.g. a more resistive background and / or a more conductive casing) increase the secondary response, as does decreasing the depth of the flaw. If only a portion of the circumference is impaired, leaving a conductive pathway connecting the top and bottom portions of the casing, the secondary signal is small and thus will be challenging to detect under most circumstances. For the subset of scenarios where we do have data sensitivity to the flaw, an inverse problem can be solved to  estimate the depth of the impairment. One approach would be to use a reduced modeling procedure whereby only a few parameters are sought. For the case presented here, we might invert for a smooth background, the length of the well, and potentially the conductivity of the casing, if it is not known a-priori.

For situations where the aim is to image a target at depth, we showed that a downhole electrode is preferable to a top-casing source for two reasons. First, for long wells, the magnitude of currents decay with distance from the source, thus having the source near the target increases the current density available to excite a response. Second, the strength of the primary at the surface is reduced if the source is downhole; this makes the secondary field a larger percentage of the primary.   For targets in close proximity to the well, if the target is in contact with the well, that electrical connection enhances the response. Our numerical modelling demonstrated that there is a non-intuitive asymmetry between the excitation of conductive targets and a resistive targets that are in contact with the well. Conductive targets have a positive charge build-up on all interfaces while resistive targets have an accumulation of both positive and negative charges. Thus, under these circumstances, conductive targets are easier to detect than resistive targets.

Finally, we considered common strategies for approximating a hollow steel-cased well and demonstrated that the product of the conductivity and the cross-sectional area of the casing is the important quantity to conserve for DC simulations. This approximation is suitable for simulation grids whose cell-widths are similar in scale to the diameter of the casing. If cell widths exceed the diameter of the casing, then more advanced numerical approaches, such as those presented in \cite{Weiss2017, Caudillo-Mata2017a} could be considered to improve accuracy.

The next set of research questions include developing strategies for solving the DC inverse problem in settings with steel-cased wells as well as extensions to time and frequency domain electromagnetics . Time-varying fields introduce inductive processes and require that magnetic permeability be considered. These factors further complicate the physics, but they also provide richer information content in the data and that will be  valuable for solving the inverse problem. In those problems, just as shown here for DC, detailed numerical simulations will continue to be a critical component for developing an understanding of the physics.

\begin{acknowledgments}

The authors are grateful to the SimPEG contributors, and in particular, Dr. Rowan Cockett and Dr. Seogi Kang for the effort they have invested in improving SimPEG. We also thank Dr. Puzyrev and the two anonymous reviewers whose critiques and feedback improved the quality of the manuscript.

The funding for this work is provided through the Vanier Canada Graduate Scholarships Program.

\end{acknowledgments}

\bibliographystyle{gji}
\bibliography{refs.bib}
\clearpage

\section{Tables}
\begin{table}
\centering
    \caption{Integrated secondary charge over a target adjacent to the casing, as shown in Figure \ref{fig:target_physics}.}
    \begin{tabular}[htb]{| r | r | r |}
        \hline
                                           & \multicolumn{2}{|c|}{\textbf{integrated secondary charge (C)}} \\
        \textbf{target conductivity (S/m)} & \textbf{downhole source} & \textbf{top-casing source} \\
        \hline
        1e-03 & -4.24e-12 & -1.08e-12 \\
        1e-02 & -3.82e-12 & -9.68e-13 \\
        1e-01 & 0.00e+00 & 0.00e+00 \\
        1e+00 & 1.75e-11 & 4.46e-12 \\
        1e+01 & 3.26e-11 & 8.28e-12 \\
        \hline
    \end{tabular}
    \label{tab:target_charge}
 \end{table}
\begin{table}
\centering
    \caption{Integrated secondary charge over a target that is not electrically connected to the casing, as shown in Figure \ref{fig:offset_target_physics}.}
    \begin{tabular}[htb]{| r | r | r |}
        \hline
                                           & \multicolumn{2}{|c|}{\textbf{integrated secondary charge (C)}} \\
        \textbf{target conductivity (S/m)} & \textbf{downhole source} & \textbf{top-casing source} \\
        \hline
        1e-03 & -4.24e-12 & -1.08e-12 \\
        1e-02 & -3.80e-12 & -9.64e-13 \\
        1e-01 & 0.00e+00 & 0.00e+00 \\
        1e+00 & 1.49e-11 & 3.79e-12 \\
        1e+01 & 2.51e-11 & 6.39e-12 \\
        \hline
    \end{tabular}
    \label{tab:offset_charge}
 \end{table}

\begin{table}
\centering
    \caption{
        Mesh parameters and computation time for each forward simulation on a 2.7 GHz Intel Core i7 processor.
        The hollow steel-cased well is discretized on the 3D cylindrical mesh with 4 cells across the thickness
        of the casing. We treat this as the baseline solution. For each of the carestian simulations, the casing is
        captured by single cell in the horizontal dimensions.
    }
    \begin{tabular}[htb]{| l | r | r | r |}
        \hline
        \textbf{Mesh Type} & \textbf{Smallest cell width (m)} & \textbf{Number of cells} & \textbf{Compute Time (s)} \\
        \hline
        3D Cylindrical & $2.5 \times 10^{-3}$ & 430528 & 9 \\
        \hline
        \multirow{4}{*}{Cartesian} & 0.1 & 1090026 & 131 \\
                                  & 0.2 & 971022 & 119 \\
                                  & 0.4 & 889998 & 87 \\
                                  & 0.8 & 738078 & 56 \\
        \hline
    \end{tabular}
    \label{tab:cartesian_simulation}
 \end{table}
\clearpage

\section{Figures}

\end{document}